\documentclass[fleqn,usenatbib]{mnras}
\usepackage[T1]{fontenc}
\usepackage{ae,aecompl}
\usepackage{amsmath}	
\usepackage{amssymb}	
\usepackage{epsfig}
\usepackage{color}
\usepackage{breakurl}
\usepackage{hyperref}

{\newif\ifnotend
\notendtrue
\def\veclist{ABCDEFGHIJKLMNOPQRSTUVWXYZabcdefghijklmnopqrstuvwxyz.}
\def\top#1#2.{#1}
\def\tail#1#2.{#2.}
\loop\expandafter\xdef\csname v\expandafter\top\veclist\endcsname%
{{\noexpand\bf\expandafter\top\veclist}}
\edef\veclist{\expandafter\tail\veclist}
\if\veclist.\notendfalse\fi\ifnotend\repeat}

\renewcommand*{\vec}[1] {\boldsymbol{#1}}
\newcommand* {\diff}    {\mathrm{d}}
\newcommand* {\sub}[2]  {{#1}_{\mathrm{#2}}}
\newcommand* {\vlos}    {\sub{v}{los}}
\newcommand* {\kms}     {\,\mathrm{km}\,\mathrm{s}^{-1}}
\newcommand* {\magn}    {\,\mathrm{mag}}
\newcommand* {\yr}      {\,\mathrm{yr}}
\newcommand* {\Gyr}     {\,\mathrm{Gyr}}
\newcommand* {\pc}      {\,\mathrm{pc}}
\newcommand* {\kpc}     {\,\mathrm{kpc}}
\newcommand* {\gl}      {{\ell}}
\newcommand* {\gb}      {{b}}
\newcommand* {\mul}     {\mu_\gl}
\newcommand* {\mub}     {\mu_\gb}
\newcommand* {\mura}    {\mu_{\mathrm{RA}}}
\newcommand* {\mudec}   {\mu_{\mathrm{DEC}}}
\newcommand* {\Var}     {\mathrm{Var}}
\newcommand* {\Cov}     {\mathrm{C}}
\newcommand* {\Vc}      {V_c}
\newcommand* {\Rg}      {R_{\mathrm{g}}}
\newcommand* {\gammavw} {\gamma_{VW}}
\newcommand* {\cross}	{\times}
\newcommand* {\RA}{{\rm RA}}
\newcommand* {\DEC}{{\rm DEC}}
\newcommand* {\mas}{\,{\rm mas}}
\newcommand* {\siglosrave}{\sub{\sigma}{los, RAVE}}

\newcommand* {\Rsun}{{R_{0}}}

\newcommand* {\Usun}{{U_{\!\odot}}}
\newcommand* {\Wsun}{{W_{\!\odot}}}
\newcommand* {\Vsun}{{V_{\!\odot}}}
\newcommand* {\Vphi}{{V_{\phi}}}
\newcommand* {\meanW}{{\overline{W}}}
\newcommand* {\Lz}{L_z} 
\newcommand* {\Tuu}{\sub{\mathsf{T}}{uu}}
\newcommand* {\Tvv}{\sub{\mathsf{T}}{vv}}
\newcommand* {\Tvw}{\sub{\mathsf{T}}{vw}}
\newcommand* {\Twv}{\sub{\mathsf{T}}{wv}}
\newcommand* {\Tuv}{\sub{\mathsf{T}}{uv}}
\newcommand* {\Tuw}{\sub{\mathsf{T}}{uw}}
\newcommand* {\Tww}{\sub{\mathsf{T}}{ww}}
\newcommand* {\Tij}{\sub{\mathsf{T}}{ij}}
\newcommand* {\Tkl}{\sub{\mathsf{T}}{kl}}
\newcommand* {\orderof}{{\mathcal{O}}}

\newcommand{\changed}[1] {{#1}}

\title[Warp, Waves, and Wrinkles in the Milky Way]{Warp, Waves, and Wrinkles in the Milky Way}
\author[R. Sch\"onrich \& W. Dehnen]{Ralph Sch\"onrich\thanks{E-mail: ralph.schoenrich@physics.ox.ac.uk} 
and Walter Dehnen\thanks{E-mail: walter.dehnen@leicester.ac.uk}
\\
Rudolf Peierls Centre for Theoretical Physics, 1 Keble Road, Oxford, OX1 3NP, UK
\\
University of Leicester, Dept. Physics \& Astronomy, University Road, Leicester LE1 7RH, UK}

\date{Draft, \today}
\pubyear{2018}

\begin{document}
\label{firstpage}
\pagerange{\pageref{firstpage}--\pageref{lastpage}}
\maketitle

\begin{abstract}
We derive unbiased distance estimates for the Gaia-TGAS dataset by correcting for the bias due to the distance dependence of the selection function, which we measure directly from the data. From these distances and proper motions, we estimate the vertical and azimuthal velocities, $W$ and $\Vphi$, and angular momentum $\Lz$ for stars in the Galactic centre and anti-centre directions. The resulting mean vertical motion $\meanW$ shows a linear increase with both $\Vphi$ and $\Lz$ at $10 \sigma$ significance. Such a trend is expected from and consistent with the known Galactic warp. This signal extends to stars with guiding centre radii $\Rg<R_0$, placing the onset of the warp at $R\lesssim7\kpc$. At equally high significance, we detect a previously unknown wave-like pattern of $\meanW$ over guiding centre $\Rg$ with amplitude $\sim1\kms$ and wavelength $\sim2.5\kpc$. This pattern is present in both the centre and anti-centre directions, consistent with a winding (corrugated) warp or bending wave, likely related to known features in the outer disc (TriAnd and Monoceros over-densities), and may be caused by the interaction with the Sgr dwarf galaxy $\sim1\Gyr$ ago. The only significant deviation from this simple fit is a stream-like feature near $\Rg\sim9\kpc$ ($|\Lz|\sim2150\kpc\kms$).
\end{abstract}


\section{Introduction}
The majority of spiral galaxies have at least some warps in their outer gas disc \citep{Bosma78,Bosma81}. Similarly, the Milky Way has a well-known warp both in its outer \ion{H}{i} \citep{Burke57, Kerr57, Weaver74} and stellar disc \citep{Djorgovski89}. Several studies also indicate that the outer disc structure is likely more complicated than a simple warp. Most point to a wave-like pattern, which can reach amplitudes in excess of $1 \kpc$ \citep[e.g.][]{Xu15,Price-Whelan15}. 

Where the Galactic warp starts is still debated. \cite{Reyle09}, fitting stellar profiles to 2MASS \citep{2MASS06} data via the Besan{\c{c}}on model \citep{Besmodel03}, place the onset of the Galactic warp at or outside the Solar annulus, whereas \cite{Drimmel01} find the onset of the warp about $1 \kpc$ inside the Solar annulus, formally excluding an onset at or outside the Solar position. \cite{Dehnen98} found the signal of a warp starting outside the Solar annulus from the velocities of Hipparcos stars; however, given the comparably small number of stars, the feature was barely significant, and debated by \cite{Seabroke07}.

Theoretical studies consider galactic warps as perturbations to otherwise axially symmetric and flat discs. Starting with the pioneering work by \cite{Hunter69}, most such investigations ignore diffusion in the disc plane and model the disc as vertically thin (an exception is \citealt{Weinberg98}, who demonstrated the presence of stable oscillating modes in a thick disc). While the approach of \citeauthor{Hunter69} allows arbitrary Fourier decompositions, already they concentrated on $m=1$ warps. Based on evidence from observations and simulations these appear to be the dominant mode of excitation. Moreover, in linear perturbation theory these $m=1$ modes do not mix with other wavenumbers, in particular $m=2,3,4$ spiral waves.

Together with the above approximations, it was thus reasonable to model warp structures as rigid, tilted concentric rings. With this assumption, the model by \cite{Sparke88} produced stable warps, which fit extragalactic observations. However, \cite*{BinneyJiang98} demonstrated that this stability hinges on the assumption of a static halo potential. With a disc composed of rigid rings in a live halo, they predicted that any warp should rapidly wrap up. This lets warps evolve into an interesting and widely unexplored regime: at a radial wavelength smaller than the typical stellar epicycle the above approximations break down, which provides a rather strict lower limit for the radial wavelength of a wrapped up warp at a couple of $\kpc$, either by Landau dampening, or by stabilisation of the pitch angle. Shorter wavelengths in the gas should (at least in the inner disc) be damped out by the interaction with the stars; in addition, turbulence in the gas disc will provide a separate lower limit for the wavelength.

The origins of the Galactic warp have been strongly debated. Most options rely on an external torque on the disc, which affects the outskirts more, yielding an increasing tilt with the decreasing moment of inertia at larger radii. Three major candidates for such external torques have been identified early on: accretion of fresh gas with misaligned angular momentum \citep[][]{Ostriker89, Quinn92, Jiang99, Roskar10}, a torque exerted by a triaxial dark-matter halo, which can also tilt the disc plane \citep[][]{Aumer13, Debattista13}, or the impact or passage by massive satellite galaxies. For the Milky Way, the most likely candidates to create and sustain a warped outer disc are the Magellanic clouds \citep[][]{Weinberg95, Weinberg06, Laporte18}, or the Sagittarius dwarf galaxy \citep[][]{Ibata98}. It should be noted that the vertical tilt of the Galactic stellar warp is likely detectable with Gaia \citep[][]{Earp17}.
Some divergent studies attempted to explain warps and vertical oscillations also by stochastic \changed{interactions of the disc with halo over-densities or subhaloes} \citep[][]{Weinberg93, Chequers17}, or interestingly by spiral patterns non-linearly coupling with and driving warp waves near their outer Lindblad resonance \citep[][]{MassetTagger97}.

A more recent line of research has been opened by the discovery of the Monoceros Ring \citep[][]{Newberg02} and the TriAnd streams \citep[][]{Majewski04}, ring-like overdensities at a bit more than twice the Solar Galactocentric radius near the outer Milky Way disc plane covering around $170\degr$ in longitude. Some earlier studies \citep[e.g.][]{Penarrubia05} quite successfully matched these observations, in particular the presence of a kinematically warm structure streaming roughly at the circular speed \citep[][]{Yanny03}, with an accretion event near the disc plane. More recent studies, however, tend to favour a picture of bending waves\footnote{Bending waves are \changed{(essentially undamped) vertical oscillations} of the disc, while breathing modes (not considered here) are symmetric around $z=0$ in density and anti-symmetric in vertical velocity with vanishing mean \citep[][]{Weinberg91, Widrow14}.} in the disc created by impacts. \cite{Kazantzidis08, Kazantzidis09} demonstrated that in addition to the known effects of minor mergers, namely warp and vertical disc heating, their simulated host galaxies produced dominant, very tightly wound spiral arms of high-density contrast, resembling ring-like structures like the Monoceros stream. Later, \cite{Gomez13} showed vertical excitations wrapping up into a spiral, very similar to the earlier predictions by \cite*{BinneyJiang98}. \cite{Onghia16} found similar structures formed within a Gyr of a Sagittarius-like impact, in particular a tightly wound spiral pattern of vertical density oscillations and corresponding mean vertical velocity oscillations with a radial wavelength of a few $\kpc$. On the observational side, \cite{Xu15} developed a comprehensive picture, providing evidence for vertical oscillations of the disc mid-plane inwards of Monoceros and TriAnd, while far above the disc, these features rotate fast and have velocity distributions as expected for disc stars.

The goal of this study is to use the first Gaia-TGAS data release \citep{GaiaDR1b, GaiaDR1, Lindegren16} to investigate the structural characteristics of the Galactic warp, and its extent in the disc. Compared to the large-scale warp, the Gaia-TGAS sample with its depth of $\sim1\kpc$ probes a small region. However, stellar kinematics will allow us to extend the reach of our analysis by the size of the stellar epicycles to $\sim3\kpc$ for the guiding centre radius in either direction from the Sun. In the stellar kinematics, the warp imprints a very small but systematic variation of the mean vertical velocity $\meanW$ with angular momentum $\Lz$, and hence the azimuthal velocity $\Vphi$. \changed{Conversely}, the measurement of a warp signal \changed{from star counts is much more} prone to systematics in the selection function: reddening is systematically biased by the position of the Sun above the plane and the patchy ISM \changed{making the southern galactic hemisphere more reddened on average and thus breaking the symmetry in star counts between north and south, which mimic a warp}. \changed{However, when kinematics and not star counts are used, this problem} can only affect the \changed{population} mixture in the sample. This can alter, but not create a warp signal in velocity space. In Gaia-TGAS, we have only 5D phase space information, since the line-of-sight velocities from Gaia are not yet published; the largest data set combining TGAS astrometry with line-of-sight velocity information, RAVE DR5 
\citep[][]{Kunder17}, has significantly fewer stars when we apply the necessary quality criteria. Hence, for most of this paper, we must do without radial velocities. The required correlation between $W$ and $\Vphi$ can only be reliably inferred in relatively narrow cones towards the Galactic centre and anti-centre. Furthermore, distance errors may lead to correlations between the inferred $W$ and $\Vphi$, requiring careful control.

This paper is structured accordingly. In Section~\ref{sec:Data} we define our coordinate system and describe the data sets we use. Distances for RAVE-TGAS have been derived by \citeauthor{SA17} (\citeyear{SA17}, hereafter SA17), but the selection function in Gaia DR1 differs significantly from RAVE-TGAS. Consequently, we derive new distances for the full Gaia-TGAS catalogue in Section~\ref{sec:Distances}. These distances are publicly available. Section~\ref{sec:theory} describes the measurement of the kinematics demonstrates the reliability of the detected warp signal. In Section~\ref{sec:measure}, we map the dependence of the mean vertical motion on Galactocentric radius, angular momentum and azimuthal velocity. Finally, in Section~\ref{sec:models} we discuss the relationship to previous conceptions of the warp and summarize our conclusions in Section~\ref{sec:Conclusions}.

\section{Data and definitions}\label{sec:Data}

\subsection{Coordinate frame and definitions}\label{sec:definitions}

Galactic longitude and latitude are termed $\gl$ and $\gb$ with proper motions $\mul=\cos\gb\,\dot{\gl}$ and $\mub=\dot{b}$, while stellar distances are denoted by $s$, the line-of-sight velocity $\vlos=\dot{s}$, observed parallax $p_0$ with uncertainty $\sigma_p$, and the actual parallax $p\equiv(\mathrm{pc}/s)\mathrm{arcsec}$. We use the standard right-handed Cartesian velocity vector ($U,\,V,\,W$), which at the Solar position coincides with the velocities in directions to the Galactic centre, Galactic rotation, and North Galactic pole, respectively. We also use standard Galacto-centric cylindrical coordinates $R$, $z$, $\phi$ and velocities $V_R=\dot{R}$, $\Vphi=R\dot{\phi}$, $W=\dot{z}$ (while $V_R=U$ and $\Vphi=V$ only for stars at $\gl=0$). We define the guiding centre as $\Rg=R\Vphi/\Vc$, using the fact that the rotation curve of the Milky Way is nearly flat in the region of interest \citep[][]{McMillan11, Piffl14}. See also Fig.~2 of SA17 for a graphical representation of these coordinates (the Galactic azimuth $\phi$ coincides with the angle $\alpha$ in that figure). $\Lz$ \changed{is positive} towards the Galactic South Pole, i.e. is negative for the disc. For the Solar motion and Galactocentric distance to the Sun, we use the determinations from \cite{S12}, \cite{McMillan11}, and \cite*{S10}, also in concordance with \cite{Reid04} and \cite{Reid14}. We set the Solar Galacto-centric distance $\Rsun=8.27 \kpc$, its total azimuthal velocity $V_{\phi,\odot}=250\kms$, and motion with respect to the Local Standard of Rest $(\Usun, \Vsun, \Wsun) = (13, 12.24, 7.24)\kms$, implying a local circular velocity of $\Vc = 238 \kms$. 
Our analysis has to concentrate on cones around the Galactic centre and anticentre directions. We define these cones by an acceptance angle $\epsilon$, where $|b| < \epsilon$ and $|l| < \epsilon$ or $|180\degr - l| < \epsilon$. To denote Galactocentric vs. heliocentric stellar positions and velocities, we use large vs. small letters, i.e. heliocentric position $\vec{r} = \vec{R} - \vec{R}_\odot$ and velocity $\vec{v} = \vec{V} - \vec{V}_{\!\odot}$.
We further split the stellar velocity into the mean at $\vec{R}$ and the peculiar motion of the star,
\begin{equation}
	\vec{V} = \bar{\vec{V}}(\vec{R}) + \vec{V}_{\!\rm p},
\end{equation}
such that the expectation value for $\vec{V}_{\!\mathrm{p}}$ (at given $\vec{R}$) vanishes. Analogously, $\bar{\vec{v}}$ and $\vec{v}_{\mathrm{p}}$ define for each star the expected and peculiar motion\changed{, respectively,} in the heliocentric frame. Here, we use the simple model of a flat rotation curve, i.e.\
\begin{equation}
	\bar{\vec{V}}(\vec{R}) = -V_{\!\mathrm{c}}\,\hat{e}_{\phi} = (Y,-X,0)^t\,V_{\!\mathrm{c}}/\sqrt{X^2+Y^2}.
\end{equation}

We split observable Heliocentric velocity $\vec{v}$ into the components parallel and perpendicular to $\vec{r}$,
\begin{equation}
	\vec{v} = \vec{p} + \vec{u},
	\quad\text{with}\quad
	\vec{u} \equiv \vlos \hat{\vec{r}} = \hat{\vec{r}}\hat{\vec{r}}\cdot\vec{v}
	\quad\text{and}\quad	
	\vec{p} \equiv \mathbfss{T}\cdot \vec{v}.
\end{equation}

In the absence of $\vlos$ information, only the transverse velocity $\vec{p}$ of the star is known, but not $\vec{u}$. We also evaluate effects of the different estimators on the measured angular momentum component $\Lz = R \Vphi = XV_y - YV_x$.

\changed{Note that a different choice of the parameters $\Rsun$, $V_{\phi,\odot}$, and $\Wsun$ would proportionally shift the values of $\Lz$, $V_{\phi}$, and $W$ derived for all stars, but hardly affect the structures detected in our study or its conclusions. Similarly, we expect that the Sun's vertical offset from the Galactic mid-plane $|z_{\odot}|$ cannot lead to a contamination of the bending-wave signal with possible breathing modes, since their scale ($\sim$ disc scale height) is much larger than $|z_{\odot}|$.}

\subsection{Data}\label{sec:ddata}
We use two different subsets: for our main analysis the full Gaia-TGAS catalogue, which has good parallaxes from Gaia DR1 
for most stars from the original Tycho catalogue, but no line-of-sight velocities, and for control the RAVE-TGAS sample, which has a smaller sample size, but combines RAVE line-of-sight velocity measurements with the parallax and proper motion information from Gaia DR1, and hence allows for full velocity measurements. We take the distances for the RAVE DR5 sample from SA17, while those for Gaia-TGAS are determined in Section~\ref{sec:Distances}.

\subsection{RAVE}\label{sec:RAVE}
A full description of the RAVE sample is provided in SA17, where we detail our quality cuts, removal of multiple entries, etc. The total RAVE DR5 sample contains 520\,701 entries. However, half the entries drop out by cross-matching with TGAS, and by excluding cluster members as well as stars with problematic line-of-sight velocity measurements ($\siglosrave > 5 \kms$, \changed{ no reasonable measurement [indicated by $\siglosrave = 0$], or} $|\vlos| > 500 \kms$). Demanding parallaxes better than $p_0/\sigma_p=5$, only 88\,516 stars remain. We note that RAVE DR5 contains further information on abnormalities in the stellar spectra, which is communicated via the flags from \cite{Matijevic12}. Those flags mark if there are particular stars among the $20$ closest matches to each RAVE spectrum. We call stars with all ``normal'' flags ``unflagged''. The ``flagged'' sample contains in particular the identified binaries.

\subsection{TGAS data}\label{sec:TGAS}
The TGAS sample from Gaia DR1 contains more than $2$ million stars. However, we require good and bias free distance determinations. Hence, we use the condition $p_0/\sigma_p>5$, for which we have validated our distance method in SA17. There are no direct kinematic selections in the TGAS sample, but there is a bias against stars with large proper motions. According to \cite{Lindegren16} TGAS loses the majority of stars with proper motions $\mu > 3.5 \arcsec\yr^{-1}$, which corresponds to a transverse velocity of $v_\perp \sim 500 \kms$ at $s = 30 \pc$. However, some objects are likely lost at smaller proper motions. In SA17, we had to exclude small distances, since a loss of high-proper motion stars could feign distance underestimates. However, the correlation between vertical and azimuthal mean velocity, which we discuss in this paper, should not be strongly impacted by a loss of high proper motion stars. The correlation between $\Vphi$ and $W$ can only be distorted if the lost stars themselves had a significant correlation (e.g. by losing stars preferentially when they are moving along a direction that is significantly inclined against $\mul$ and $\mub$). To test this, we measured the correlation between $\Vphi$ and $W$ when varying a distance cut excluding nearby stars, i.e. $s > s_{\mathrm{inner}}$. For values between $s_{\mathrm{inner}} = 0$ and $100 \pc$ there was no significant change in our findings.

\begin{figure}
    \epsfig{file=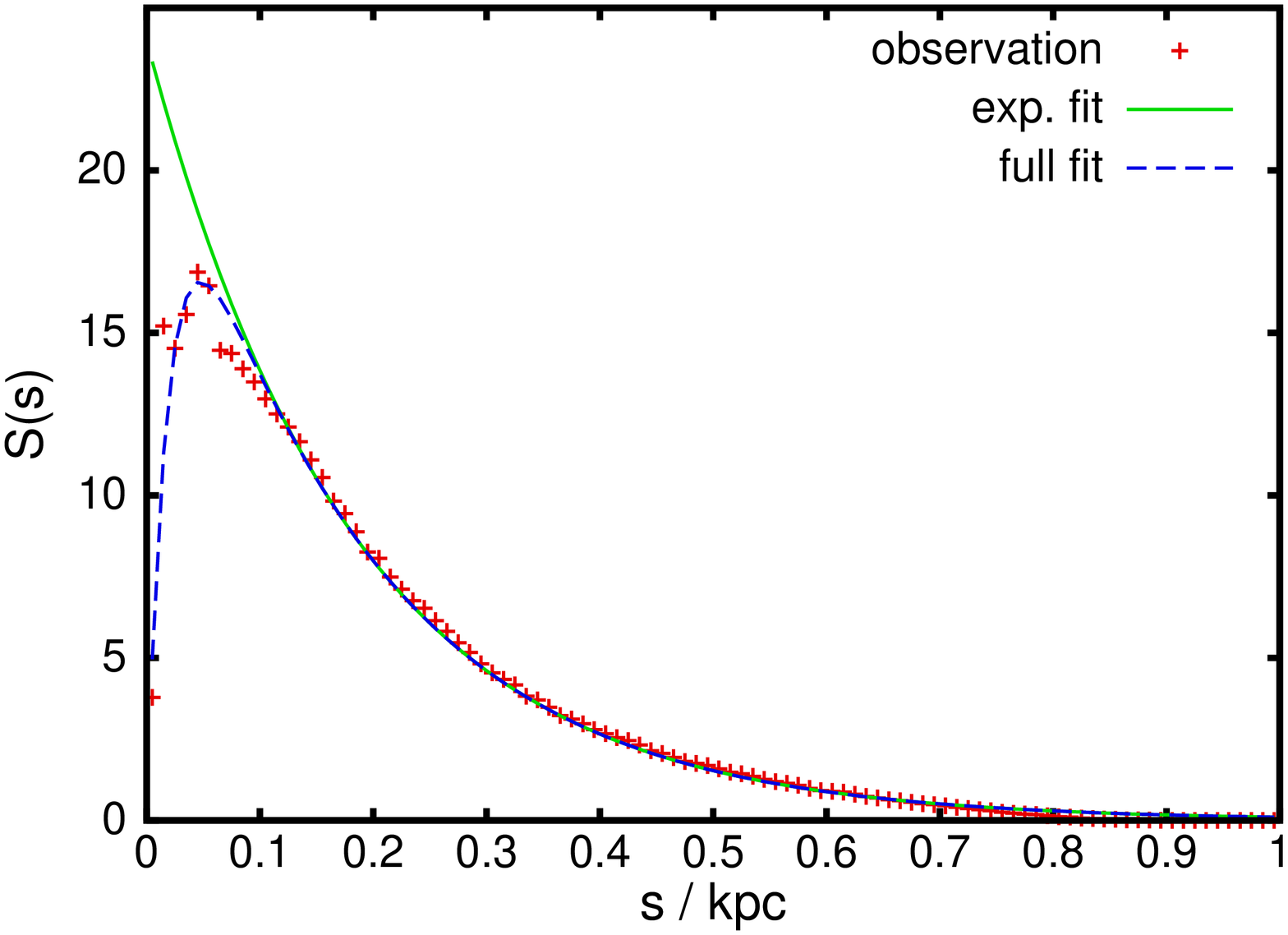,angle=-90,width=0.99\hsize}\\
    \epsfig{file=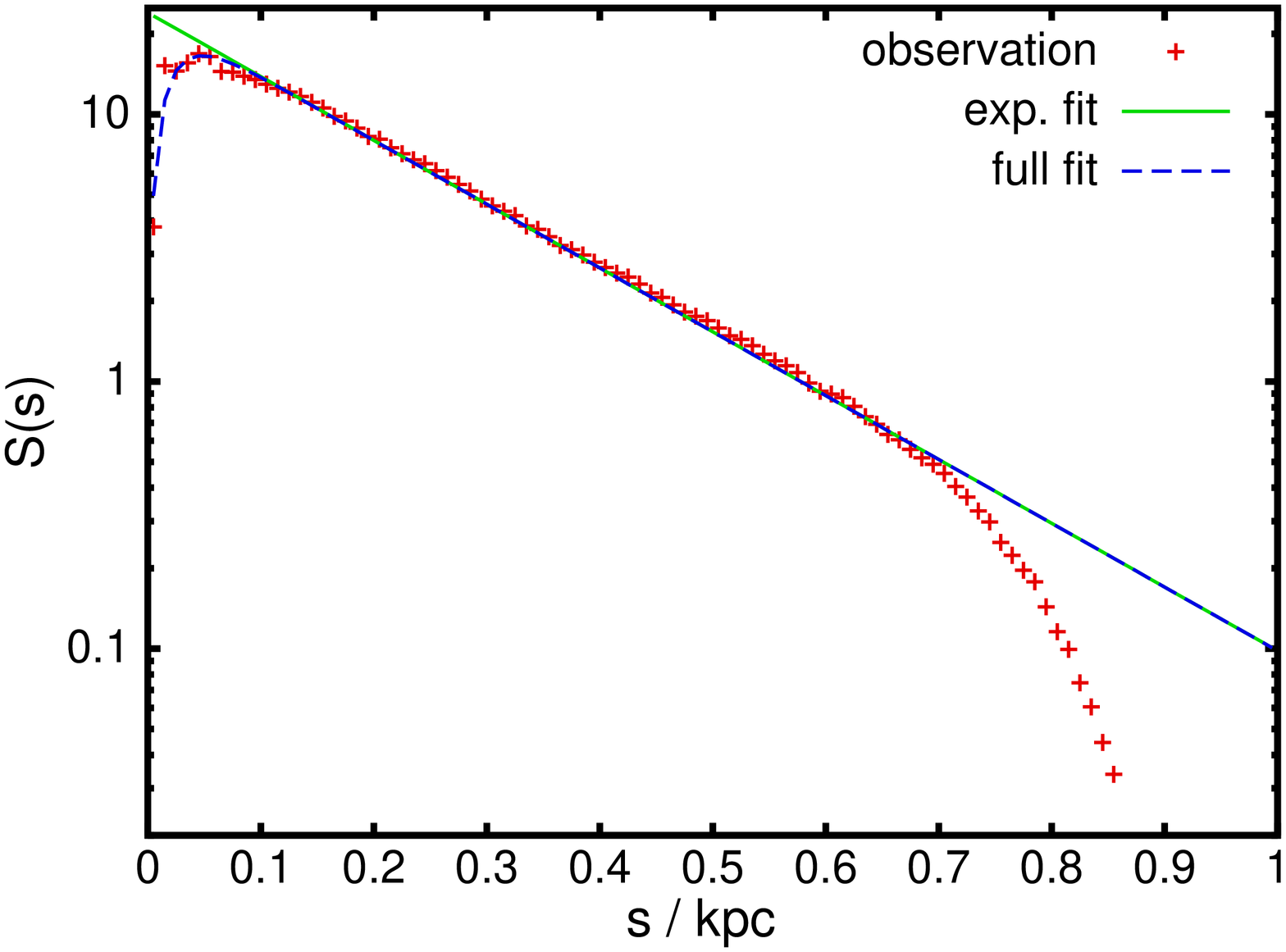,angle=-90,width=0.99\hsize}
\caption{Deriving the selection function $S(s)$ in distance. Crosses show the ratio between observed stellar density, $\sub{\rho}{obs}(s)$, and the expected density $\rho(s)$ from equation (\ref{eq:rho}). Dashed and solid curves show the analytic fit~\eqref{eq:sel} to $S(s)$ and its dominant exponential decline, respectively. The top plot uses a linear, the bottom plot a logarithmic scale for $S(s)$.}\label{fig:priortest}
\end{figure}

\section{Distance Determinations and Selection Function for Gaia DR1}\label{sec:Distances}

For retrieving distances to single stars, we use the method of SA17. Since we are dealing with first-order moments of the velocities, which are linear in the distance given the measured proper motions, we require expectation-true distance estimates, i.e.\ we need to find the right priors for our Bayesian distance estimator. For a discussion on deriving distances from parallaxes, we refer to \cite{Stromberg27}, \cite{SB14}, and \cite{Astraatmadja16}. Like SA17, we do not model the stellar photometry, and hence have to write the colour-magnitude based selection function $S$ into the priors. We assume that this selection function is isotropic, i.e. depends only on distance. We use equation~(1) from SA17:
\begin{equation}\label{eq:distex}
    \left<s\right> = N^{-1} \int \mathrm{d}s \, s^3\, G\left(p(s)|p_0, \sigma_{p}\right) \,\rho(s,\gl,\gb) \,S(s),
\end{equation}
with normalisation
\begin{equation}
    N =  \int \mathrm{d}s \, s^2\, G\left(p(s)|p_0, \sigma_{p}\right)\, \rho(s,\gl,\gb)\,S(s).
\end{equation}
Here, $G(p|p_0, \sigma_{p})$ is the observational likelihood of the parallax $p$ given a measurement $(p_0, \sigma_{p})$, while $\rho(s,\gl,\gb)$ is the Galactic stellar density. We employ the simple density model \citep{SB14}:
\begin{equation}\label{eq:rho}
    \rho(R,z) \propto 
    \mathrm{e}^{-\frac{R-\Rsun}{R_d}}
    \left[\mathrm{e}^{-\frac{|z|}{z_0}}
    + a_t \mathrm{e}^{-\frac{|z|}{z_{0,t}}}
    \right]
    + a_h \left(\frac{r}{\Rsun}\right)^{-2.5}
\end{equation}
with disc scale length $R_d = 2.5 \kpc$, relative thick disc normalisation $a_t = 0.12$, and disc scale heights $(z_0, z_{0,t}) = (0.3,0.9) \kpc$ in agreement with \cite{Ivezic08}. $a_h = 0.001$ normalises the (unimportant) halo component.

In SA17 we found that the selection function in distance $s$ can be reconstructed from the data by fitting an analytic function $S'(s) \sim \sub{\rho}{obs}(s)/\rho(s)$ to the ratio between the observed density of stars $\sub{\rho}{obs}(s)$ to the simple model~\eqref{eq:rho}. The final approximation \changed{for $S(s)$ is gained by starting with a flat/constant S(s). We then iterate} by re-inserting the new fit for $S(s)$ into equation~\eqref{eq:distex} until the selection function converges. For Gaia DR1, $S(s)$ is fit almost perfectly by a simple declining exponential:
\begin{equation}\label{eq:sel}
    S(s) \propto \exp{(-bs)} [1 - \exp(-b_2 s)],
\end{equation}
where $b = 5.51 \kpc^{-1}$. The second term models a decline of stars at the smallest distances with $b_2=47.5 \kpc^{-1}$ and was inserted for purely aesthetic reasons to account for the bright limit of the survey at magnitude $G \sim 6 \magn$. Neither the distance estimates nor any of our findings are significantly impacted by this term.

The resulting selection function is shown in Fig.~\ref{fig:priortest}. The analytic function provides an excellent fit. The breakdown for distances $s > 0.75 \kpc$ is caused by an increasing fraction of stars dropping out of the sample due to the parallax quality cut (which must not be written into the selection function). We also confirm that the shape of the selection function is the same in the Galactic plane ($|\gb| < 10\degr$) and towards Galactic poles ($|\gb| >70\degr$). The overall normalisation is different for stars at low latitudes, which is caused both by the assumed density model not perfectly fitting the lowest altitudes $|z|$, and by the loss of stars in the survey near the plane due to crowding and extinction. However, since the shape of $S(s)$ remains the same, this does not significantly impact the distance estimates.

The distances derived with this method are available.\footnote{Please find them at \burl{http://www-thphys.physics.ox.ac.uk/people/RalphSchoenrich/data/tgasdist/tgasdist.tar.gz}.}

\section[Estimators for $\Vphi$ and $W$]{\boldmath Estimators for $\Vphi$ and $W$}\label{sec:theory}

Readers, who are mostly interested in our measurement results are recommended to continue in Section 5 and refer to Section 4.2.1 if needed. This Section derives and validates our method for the expectation values for the azimuthal ($\Vphi$) and vertical ($W$) velocity components and their covariance for the observed stars, which is relevant for the measurement of the galactic warp signal. In the second half of the section, we will discuss three different methods to obtain less biased $\Vphi$ and $W$ velocity components and their detailed bias terms. Throughout this Section, we will indicate estimated quantities with primes. Our method to derive distances from parallaxes has been tested in SA17 to be bias-free at the detection limit of order $\sim1\%$. Thus, for most of the discussion we neglect the effects of distance bias \citep[which is best assessed with the same line of formalism, see][]{SBA}. We have further evaluated (see Appendix) that the effects of random parallax errors and error correlations with proper motions have no significant effect on our results.

\subsection{Method}\label{sec:theory:method}
In a full measurement, like RAVE-TGAS, we have proper motions in the Galactic longitude and latitude ($\mul,\,\mub$) directions, a distance estimate ($s$), and a line-of-sight velocity  ($\vlos$). The Cartesian motions in the heliocentric frame are then obtained by
\begin{equation}
    \begin{pmatrix} U_0 \\ V_0 \\ W_0 \end{pmatrix} = \mathbfss{M}\cdot
    \begin{pmatrix} s\mub \\ s\mul \\ \vlos \end{pmatrix}\quad\text{and}\quad
    \begin{pmatrix} s\mub \\ s\mul \\ \vlos \end{pmatrix} = \mathbfss{M}^{T}\cdot
    \begin{pmatrix} U_0 \\ V_0 \\ W_0 \end{pmatrix}
\end{equation}
with the orthogonal (rotation) matrix
\begin{equation}
    \mathbfss{M}\equiv\begin{pmatrix}
    -\sin\gb \cos\gl  & \,\,\, -\sin\gl \,\,\, & \cos\gb \cos\gl  \\
    -\sin\gb \sin\gl  & \phantom{-}\cos\gl  & \cos\gb \sin\gl  \\
    \cos\gb  & 0 & \sin\gb \end{pmatrix}.
\end{equation}
However, as TGAS lacks any measurement of $\vlos$, this relation cannot be used to find the full velocities of individual stars in the TGAS sample. The problem of missing $\vlos$ information becomes apparent when we set $\vlos = 0$ in our equations and relate the inferred velocity components $(U', V', W')$ to the real heliocentric velocity vector $(U_0, V_0, W_0)$:
\begin{equation}\label{eq:veltrans}
    \vec{v}' = \begin{pmatrix} U' \\ V' \\ W' \end{pmatrix} \equiv
    \mathbfss{M} \cdot \begin{pmatrix} s\mub \\ s\mul \\ 0 \end{pmatrix} = \mathbfss{T} \cdot \begin{pmatrix} U_0 \\ V_0 \\ W_0 \end{pmatrix} $,$
\end{equation}
where $\mathbfss{T} \equiv\mathbfss{M}\cdot\mathrm{diag}(1,1,0)\cdot\mathbfss{M}^t$ is the symmetric projection matrix \citep*{DehnenBinney98,SBA}. We have:
\begin{align}
\label{eq:T}
    \mathbfss{T}
    &= \mathbfss{1} - \hat{\mathbfit{r}}\hat{\mathbfit{r}}^t = \begin{pmatrix}
		\tilde{y}^2+\tilde{z}^2 & -\tilde{x}\tilde{y} & -\tilde{x}\tilde{z} \\
		-\tilde{x}\tilde{y} & 1-\tilde{y}^2	  & -\tilde{y}\tilde{z} \\
		-\tilde{x}\tilde{z} & -\tilde{y}\tilde{z} & 1-\tilde{z}^2
	\end{pmatrix} \\&= \begin{pmatrix} 
1-\cos^2\!b\cos^2\!l  &\;& -\frac12\cos^2\!b\sin 2l &\;& -\frac12\sin 2b \cos l
\\[0.5ex]
-\frac12\cos^2\!b \sin2l  && 1-\cos^2\!b \sin^2\!l && -\frac12\sin 2b \sin l 
\\[0.5ex]
-\frac12\sin 2b \cos l  && -\frac12\sin 2b \sin l && 1 - \sin^2\!b
\end{pmatrix}.
\end{align}
Here, 
\begin{equation} 
\hat{\mathbfit{r}} := (\hat{x}, \hat{y}, \hat{z})^t \equiv
	(\cos b\cos\ell,\,\cos b\sin\ell,\,\sin b)^t 
\end{equation}
 is the unit vector pointing from the Sun to the star and is identical to the last column of $\mathbfss{M}$. In a sample with $\vlos$ information, $\mathbfss{T}$ describes the correlations between observed velocities and $(\gl,\gb)$ arising from biased distances, and so opens the way to precise statistical distance tests. Here, it provides the bias caused by missing $\vlos$. We denote the components of $\mathbfss{T}$ after the velocities they connect, i.e.\ $\Tvv$ is the diagonal element for the azimuthal velocity component and $\Tvw=\Twv$ connects $V$ and $W$ velocity components. Good $V$ and $W$ estimates require
\begin{equation}
    \Tvv = 1 - \sin^2\gl \cos^2\gb \approx 1
    \quad\text{and}\quad
    \Tww = 1 - \sin^2\gb \approx 1.
\end{equation}
Obviously, this is satisfied only near the centre and anticentre directions, i.e.\ $(\gl,\gb) \sim (0\degr, 0\degr)$ or $(\gl,\gb) \sim (180\degr,0\degr)$. Near these lines (assuming angles in radian)
\begin{equation}
    \label{eq:T:taylor}
    \mathbfss{T} \sim 
    \begin{pmatrix}
    \gb^2+\bar{\gl}^2 & -\bar{\gl} & -\gb \\
    -\bar{\gl} & 1-\bar{\gl}^2 & -\bar{\gl}\gb \\
    -\gb & -\bar{\gl}\gb & 1-\gb^2
    \end{pmatrix}
\end{equation}
with $\bar{\gl}=\gl$ (centre) or $\bar{\gl}=\gl-180\degr$ (anti-centre). While the diagonal elements $\Tvv$ and $\Tww$ are quadratic in either $\bar{\gl}$ or $\bar{\gb}$, the the off-diagonal elements are linear, i.e. minimising biases will demand a sample symmetric in $\gl$ and $\gb$, in order to cancel these terms.

\subsubsection{A correct trend between $V$ and $W$ velocity}
\label{sec:basic}
Our analysis requires not only unbiased expectation values for each star's $V$ and $W$, but also the correct slope between the inferred $V'$ and $W'$, which in a linear regression is determined by their covariance $\Cov(V',W')$. Inserting equation~\eqref{eq:veltrans} into $\Cov(V',W')$, we have
\begin{eqnarray}
\Cov(V',W') &=& \Cov(\Tuv U_0 + \Tvv V_0 + \Tvw W_0,
\nonumber\\ &&\phantom{ \Cov(}
        \Tuw U_0 + \Tvw V_0 + \Tww W_0)
\nonumber\\ &=&
\nonumber
\Cov(\Tvv V_0, \Tww W_0) + \Cov(\Tvw V_0, \Tvw W_0) \quad \quad \\ \nonumber
&+& \Cov(\Tuv U_0, \Tuw U_0) + \Cov(\Tvv V_0, \Tvw V_0)
\\ \nonumber
&+& \Cov(\Tvw W_0, \Tww W_0)
\\ \nonumber
&+& \Cov(\Tuv U_0, \Tww W_0) + \Cov(\Tuw U_0, \Tvv V_0) \\ 
&+& \Cov(\Tuw U_0, \Tvw W_0) + \Cov(\Tuv U_0, \Tvw V_0). \label{eq:cov}
\end{eqnarray}
Clearly, the first two terms on the right-hand side of equation~\eqref{eq:cov} are our main targets --- the first term deviates from the desired value $\Cov(V_0,W_0)$ by an amount $\orderof(\epsilon^2)$, where $\epsilon$ is the angular distance from the Galactic centre or anti-centre directions. However, we can correct the expectation value by defining new velocity components $(U'',V'',W'')$:
\begin{equation}\label{eq:velcorr}
    \begin{pmatrix} U'' \\ V'' \\ W'' \end{pmatrix} \equiv \begin{pmatrix} U'/\Tuu \\ V'/\Tvv\\ W'/\Tww \end{pmatrix},
\end{equation}
when
\begin{eqnarray}
    \label{eq:fincov}
    \Cov(V'',W'') &=& \Cov(V_0, W_0)
    \\ \nonumber
    &+& \Cov(V_0, \Tvw/\Tww V_0) + \Cov(\Tvw/\Tvv W_0, W_0)
    \\ \nonumber
    &+& \Cov(\Tuv/\Tvv U_0, \Tuw/\Tww U_0)
    \\ \nonumber
    &+& \Cov(\Tvw/\Tww V_0, \Tvw/\Tvv W_0)
    \\ \nonumber
    &+& \Cov(\Tuv/\Tvv U_0, \Tvw/\Tww V_0)
    \\ \nonumber
    &+& \Cov(\Tuw/\Tww U_0, \Tvw/\Tvv W_0)
    \\ \nonumber
    &+& \Cov(\Tuw/\Tww U_0, V_0) + \Cov(\Tuv/\Tvv U_0, W_0).
\end{eqnarray}
The first term is exactly what we want. The remaining correction terms in equation~\eqref{eq:fincov} have very different levels of expected impact on our results. The mixed-velocity correlations in equation~\eqref{eq:fincov} are in general orders of magnitude smaller than the covariance terms with the same velocity component on both sides (second to fourth terms on the right hand side). From equation~\eqref{eq:T:taylor}, we see that these three terms are all $\orderof(\bar{\gl}\gb)$, i.e.\ they are second order in the angular offset $\epsilon$ from the centre or anti-centre directions and are both north-south and left-right anti-symmetric at these directions. At vanishing expectation values, and if the variance $\Var(y)=\Cov(y,y)$ is independent of other variables, we have $\Cov(xy, zy)\sim\Cov(x,z) \Var(y)$. That is, in this case, if terms like $\Tvw/\Tww$ have non-vanishing sample mean, they result in large deviations on the measured correlation. Owing to the anti-symmetries of $\Tvw$, $\Tuv$, and $\Tuw$ (see equation~\ref{eq:T:taylor}), this problem is diminished if we keep our selection cones strictly symmetric in $\bar{\gl}$ and $\gb$, such that these terms average out at $\orderof(\epsilon^2)$, leaving only terms $\orderof(\epsilon^3)$.

\begin{figure}
\epsfig{file=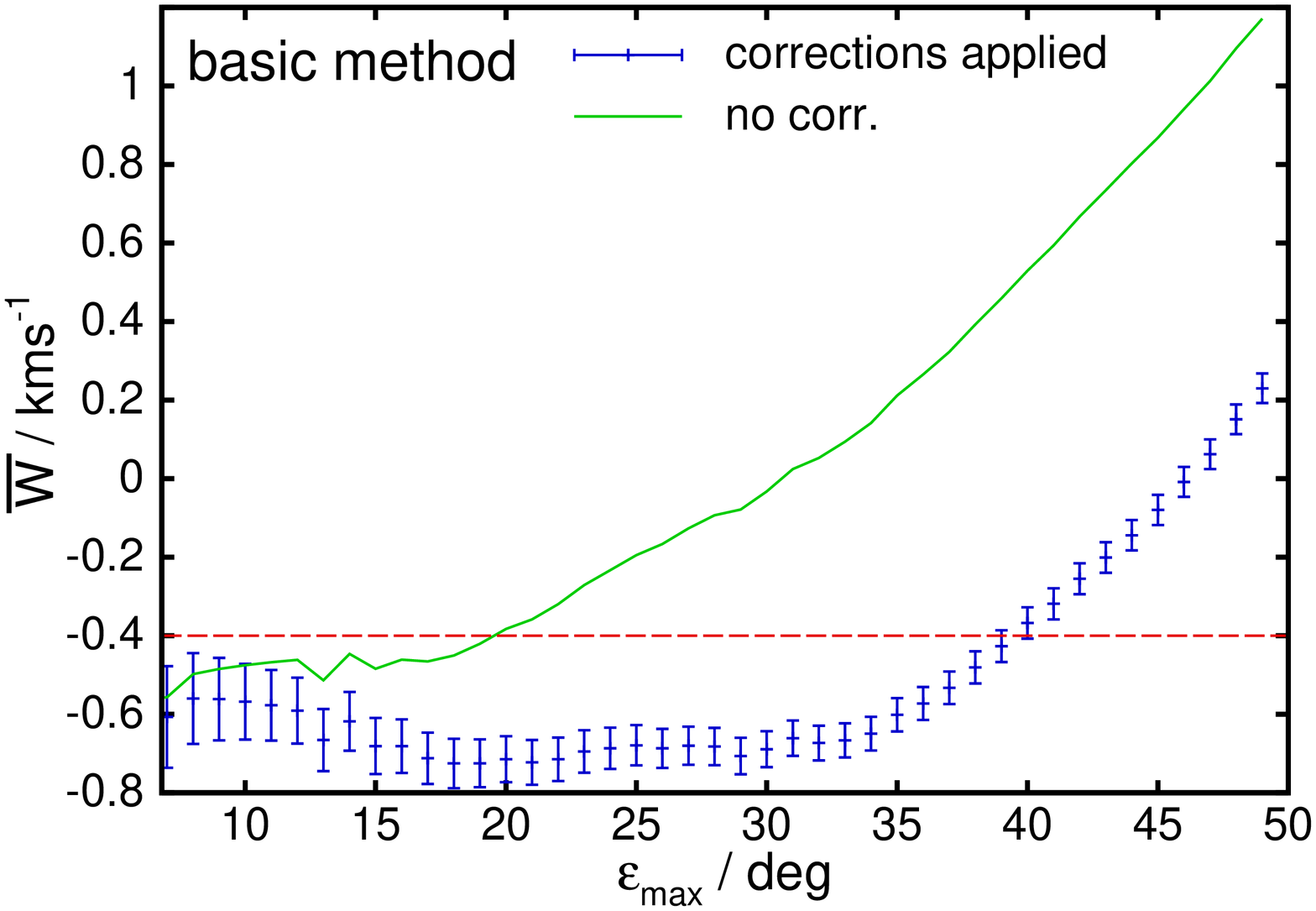,angle=-90,width=\hsize}\\
\epsfig{file=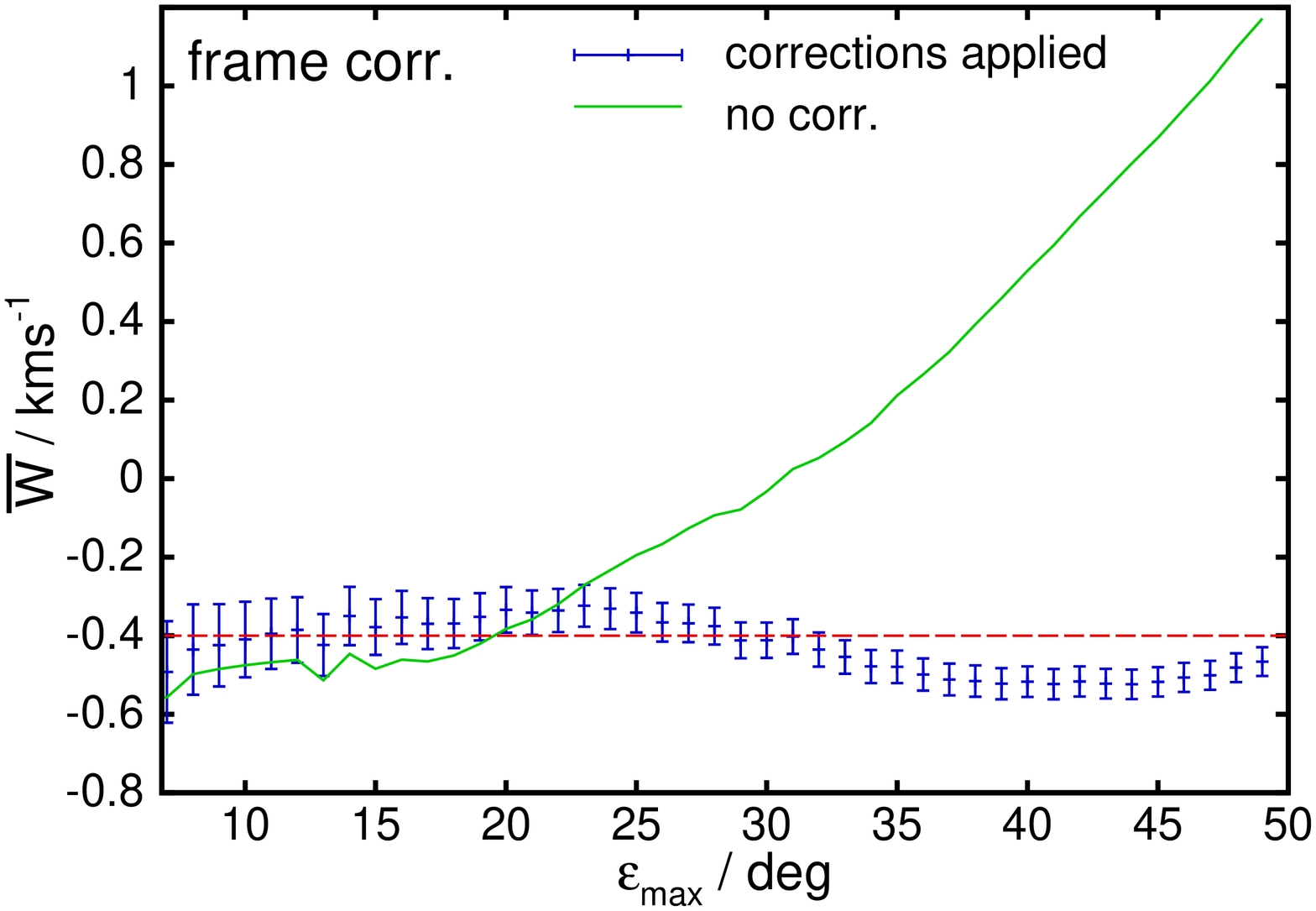,angle=-90,width=\hsize}
\caption{Effect of the velocity correction from equation~\eqref{eq:velcorr}. The error bars in the top panel show $W''$ using the basic correction, and the error bars in the bottom panel show $W''$ with a full frame correction, while the dashed line shows $W'$ \changed{(obtained without correction, i.e.\ assuming $\vlos=0$)}. The horizontal line just gives a visual aid to compare the two plots.
\changed{
The small offset in $\meanW$ between basic and frame corrected method for the entire sample is explained (see equation~\ref{equ:diag} and the last paragraph of Section~\ref{sec:error:frame:corr}) by a small contamination ($\approx 0.2 \kms$) of the basic method with the radial component of the Solar reflex motion.
}
}\label{fig:meanstat} 
\end{figure}

\subsection{Methods for velocity correction}
\subsubsection{Overview}
To control systematic errors in the basic correction method, we employ either of the three following strategies.
\begin{itemize}
 \item The \emph{basic method}, as described in Section~\ref{sec:basic} above, accounts for the missing line-of-sight velocities by dividing all heliocentric velocities by the respective diagonal elements of $\mathbfss{T}$.
 \item The \emph{frame corrected} method indirectly accounts for $\vlos$ expectations by evaluating the motion of each star in a local frame co-rotating with the disc at an azimuthal velocity $\Theta = 231 \kms$. In this frame, the mean line-of-sight velocity will be close to zero. The steps of this analysis are: i) subtract the expected mean proper motion resulting from the Solar and frame motions for each star, ii) evaluate the heliocentric cartesian velocities in that frame, iii) correct with equation~\eqref{eq:velcorr}, iv) rotate back into the Galactocentric cylindrical frame.
 \item The \emph{inversion method} relies on a local inversion of the $V$ and $W$ parts of the matrix $\mathbfss{T}$. While $\mathbfss{T}$ cannot be fully inverted due to the missing $\vlos$ information -- $\mathbfss{T}$ maps onto a two-dimensional plane in velocity space (dim(kern($\mathbfss{T}$)) = 1) -- , a partial inversion is appropriate near the centre-anticentre line. We hence obtain the velocities by dividing $U$ by $\Tuu$ and multiplying $(V,W)$ with the inverse of the corresponding $2\times2$ sub-matrix of $\mathbfss{T}$. 
\end{itemize}

We will concentrate most of our analysis on the first two methods, but show the most important statistics also for the inversion technique. A detailed discussion of biases in each method is provided in Section~\ref{sec:detailed:analysis} below,
\changed{but first we consider the naive and common approach of not correcting at all, i.e.\ using} the proper motion part only. This is equivalent to setting $\vec{u}=0$ such that
\begin{subequations}
\begin{align}
	\vec{v}'&=\vec{p} = \mathbfss{T}(\vec{V} - \vec{V}_{\!\odot})
		&&=\vec{V}-\vec{V}_{\!\odot}
		- \hat{\vec{r}}\hat{\vec{r}} \cdot \vec{v},\qquad\;
	\\
	\vec{V}'&=\vec{v}'+\vec{V}_{\!\odot}
		&&=\vec{V}
		- \hat{\vec{r}}\hat{\vec{r}}\cdot\vec{v}
	\\
	W'&&&= W - \tilde{z}\;\hat{\vec{r}}\cdot\vec{v}
	\\
	\vec{L}' &=\vec{R}\cross\vec{V}' && = \vec{L} - \Rsun (\hat{e}_x\cross\hat{\vec{r}})\vlos 
	\\
	L_z' &&&= L_z + \Rsun \hat{y} \vlos. 
\end{align}
\end{subequations}
For the angular momentum terms, we used $\vec{R}\cross\hat{\vec{r}} = (\vec{R_0} + \vec{r})\cross\hat{\vec{r}} = \Rsun (\hat{e}_x\cross\hat{\vec{r}})$ and $\hat{\vec{r}}\cdot\vec{v} = \vlos$.

Since the expectation value for $\hat{\vec{r}}\cdot\vec{v}\neq0$, this method suffers a systematic error (bias) in both $W'$ and $L_z'$, which grows linearly with angular distance $(\tilde{y},\tilde{z})$ from the centre or anti-centre directions. The error in $W'$ is not fully eliminated by north-south symmetry, as both $\tilde{z}$ and $\overline{\hat{\vec{r}}\cdot\vec{v}}$ change sign when $\gb$ does. Analogously, the bias in $L_z'$ has non-zero expectation value even with east-west symmetry in $\gl$.

We note further that \changed{accounting for the missing line-of-sight velocity information $\vec{u}$ by using its expectation instead} does not resolve \changed{our} problems, since we want to measure deviations from the mean for sub-samples in the survey: in this case $W' = W - \tilde{z}\,\hat{\vec{r}}\cdot\vec{V}_{\!\mathrm{p}}$, and similar to the above discussion, $W'$ is still biased. In short, this simple approach should be avoided.

\changed{These effects can be seen in Fig.~\ref{fig:meanstat},} which shows the dependence of $\meanW$ \changed{for} the entire sample versus the cone opening angle $\epsilon$. The solid lines demonstrate how without corrections $\meanW$ already starts drifting at small $\epsilon$. \changed{Conversely,} the stabilising effect of the corrections on the mean vertical velocity \changed{is evident from the data points with error bars} (top panel: basic method, bottom panel: frame correction).


\subsubsection{Analysis of the correction methods}
\label{sec:detailed:analysis}

We now have two points of attack, of which we present three combinations throughout the paper: 
\begin{enumerate}
 \item \label{item:corr:one} correcting the frame, i.e. move to a \changed{velocity} frame where \changed{the expected stellar velocity vanishes (}which leaves only the peculiar velocity part of the proper motion\changed{)}, and
 \item \label{item:corr:two} scaling the proper motions to account for the missing information, which amounts technically to inserting a new mapping/matrix $\tilde{\mathbfss{T}}$.
\end{enumerate}

For step \ref{item:corr:one} we first subtract the expectation
\begin{equation}
	\bar{\vec{p}} \equiv \mathbfss{T}\cdot(\bar{\vec{V}}-\vec{V}_{\!\odot})
\end{equation}
from $\vec{p}$, \changed{then in step \ref{item:corr:two}} correct for lack of line-of-sight velocity in that frame, \changed{before adding} $\bar{\vec{V}}-\vec{V}_{\!\odot}$ \changed{to convert} back \changed{to the heliocentric velocity frame}. This corresponds to the following \changed{combined} correction:
\begin{align}
	\vec{v}' &= \tilde{\mathbfss{T}}\cdot(\vec{p}-\bar{\vec{p}})
		+ \bar{\vec{V}}-\vec{V}_{\!\odot}\!\!\!\!
	&&= \tilde{\mathbfss{T}}\cdot\mathbfss{T}\cdot\vec{V}_{\!\mathrm{p}}
		+ \bar{\vec{V}}-\vec{V}_{\!\odot}
	\\
	\vec{V}' &= \vec{v}' + \vec{V}_{\!\odot}
	&&= \tilde{\mathbfss{T}}\cdot\mathbfss{T}\cdot\vec{V}_{\!\mathrm{p}}
		+ \bar{\vec{V}}
\end{align}
We now consider two options for the correction matrix $\tilde{\mathbfss{T}}$:

\paragraph{Dividing by the diagonal elements of $\mathbfss{T}$}
\label{sec:error:frame:corr}
\changed{This is the same approach as in the basic method of Section~\ref{sec:basic} (only that here we apply it in each star's standard of rest rather than the local standard of rest) and corresponds to}
\begin{subequations}
\label{eqs:frame:corr}
\begin{equation}
	\tilde{\mathbfss{T}} = 
	\begin{pmatrix}
		(\tilde{y}^2+\tilde{z}^2)^{-1} 	& 0 & 0 \\
		0 & (1-\tilde{y}^2)^{-1} & 0 \\
		0 & 0 & (1-\tilde{z}^2)^{-1}
	\end{pmatrix}
\end{equation}
such that
\begin{equation}
	\tilde{\mathbfss{T}}\cdot\mathbfss{T} =
	\begin{pmatrix}
		1 & -\tilde{x}\tilde{y}/(\tilde{y}^2+\tilde{z}^2) &
			-\tilde{x}\tilde{z}/(\tilde{y}^2+\tilde{z}^2) \\
		-\tilde{x}\tilde{y}/(1-\tilde{y}^2) & 1 &
		-\tilde{y}\tilde{z}/(1-\tilde{y}^2) \\
		-\tilde{x}\tilde{z}/(1-\tilde{z}^2) & 
	-\tilde{y}\tilde{z}/(1-\tilde{z}^2) & 1
	\end{pmatrix}.
\end{equation}
This gives \changed{for our \emph{frame-corrected} method}
\begin{align}
	\label{eq:frame:corr:W}
	W' &= W - \frac{\tilde{z}}{1-\tilde{z}^2}(\tilde{x}V_{{\rm p},x}+\tilde{y}V_{\mathrm{p},y}) \\ \label{equ:diag}
	&=W - (\beta_{U,\mathrm{diag}}V_{\mathrm{p},x}+\beta_{V,\mathrm{diag}}V_{\mathrm{p},y})\\[1ex]
	\label{eq:frame:corr:Lz}
	L_z' &= L_z
		+\frac{\tilde{x}\tilde{y}r}{\tilde{y}^2+\tilde{z}^2}(\tilde{y}V_{\mathrm{p},y}+\tilde{z}V_{\mathrm{p},z})
		-\frac{\tilde{y}X}{1-\tilde{y}^2}(\tilde{x}V_{\mathrm{p},x}+\tilde{z}V_{\mathrm{p},z}),
\end{align}
\end{subequations}
where we have replaced $Y=y=\tilde{y}r$. \changed{The $\beta_i$ coefficients are implicitly defined by these equations and can be calculated from the stellar sky positions.} Thus, the error in $W$ is harmless: not only is it linear in angular distance from the centre or anti-centre directions \changed{(such that its average over symmetric samples vanishes)}, but also only proportional to the star's peculiar motion $\vec{V}_{\!\mathrm{p}}$. The error in $L_z$ is dominated by the first error term, which is again proportional to $\vec{V}_{\!\mathrm{p}}$, but does not vanish in the limit $(\tilde{y},\tilde{z})\to0$ of the centre or anti-centre directions (in contrast to the second term, which does).

\changed{The relations for our basic method are structurally identical and can be obtained from equations~(\ref{eqs:frame:corr}c-e) by replacing the star's peculiar velocity $\vec{V}_{\!\mathrm{p}}$ with its heliocentric velocity $\vec{v}=\vec{V}-\vec{V}_{\!\odot}$.}

\begin{figure}
\epsfig{file=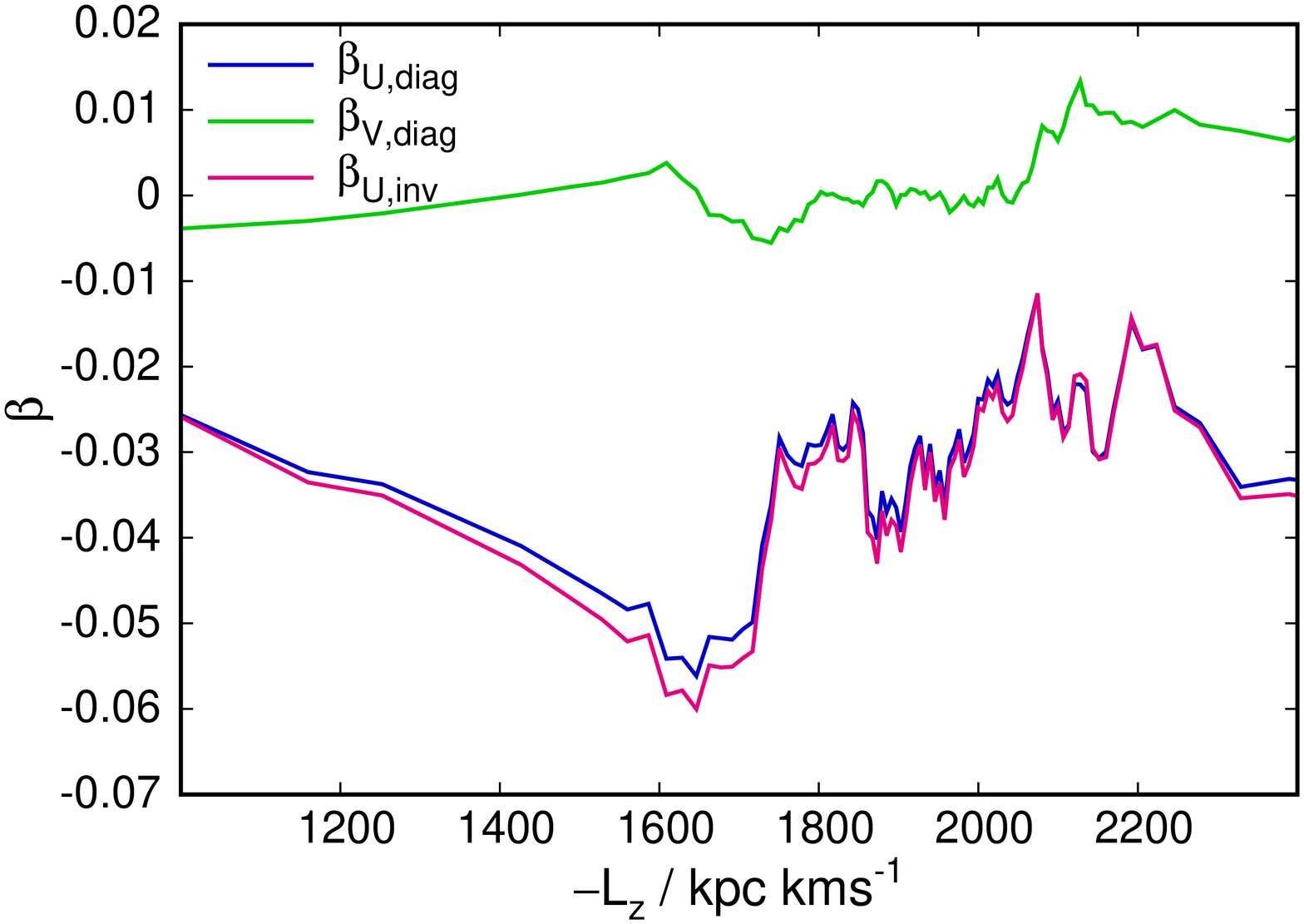,angle=-90,width=\hsize}
\epsfig{file=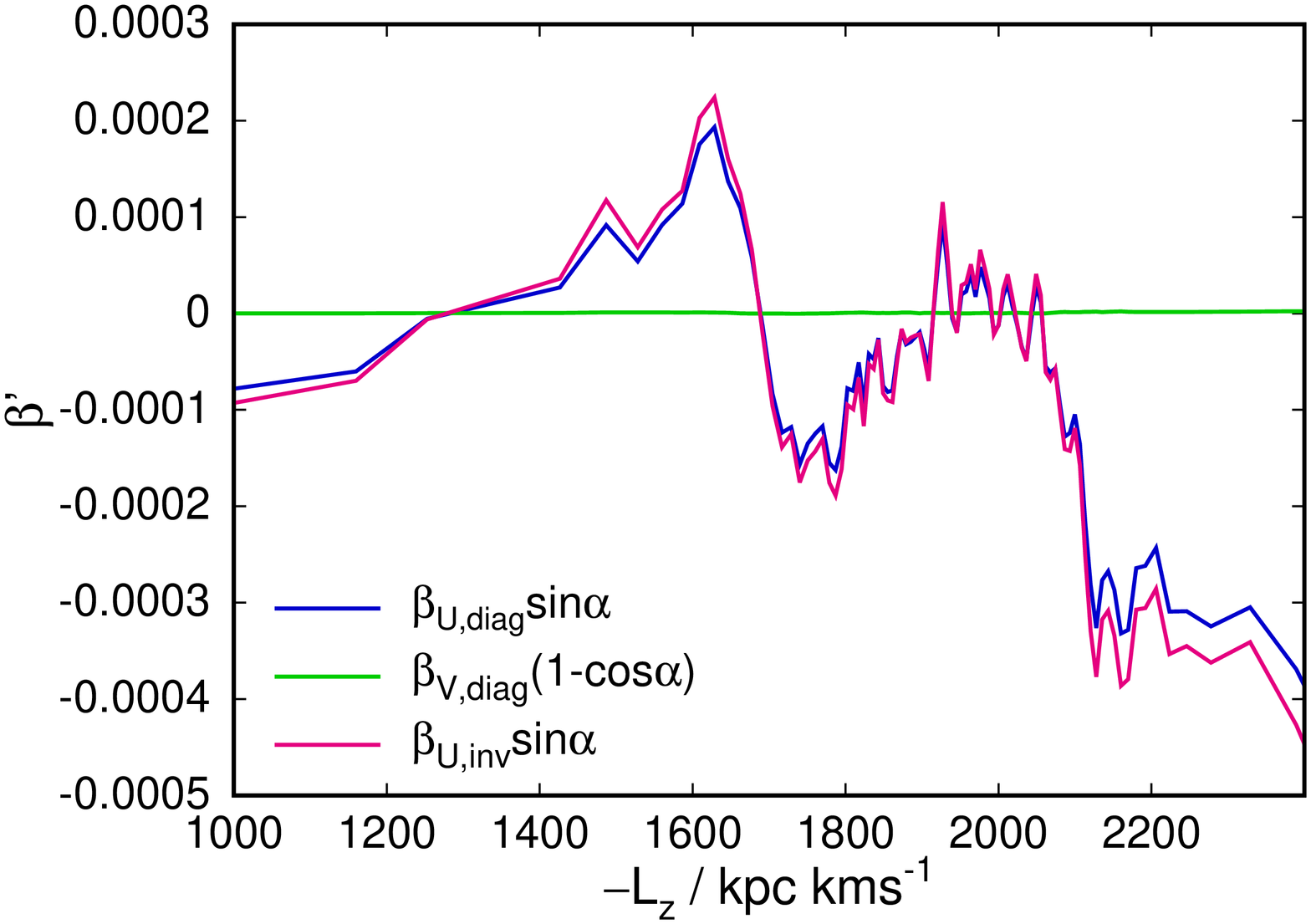,angle=-90,width=\hsize}
\caption{The connecting coefficients between the error in $W$ and the in-plane components of $\vec{V}_{\!\mathrm{p}}$ from equations \eqref{equ:diag} and \eqref{equ:inv}, for the sample separated in $\Lz$ calculated with the basic method. The top panel shows the mean values of $\beta_i$ at each bin in $\Lz$. To assess the risk from galactic rotation, the second panel displays the mean values of $\beta_{U}\sin{\alpha}$ and $\beta_{V}(1-\cos{\alpha})$.}\label{fig:coeff}
\end{figure}

\paragraph{Partially inverting $\mathbfss{T}$} \changed{The idea here is similar, but instead of merely dividing by $\Tvv$ and $\Tww$, we} invert the 2x2 sub-matrix of $\mathbfss{T}$ which relates to the $y$ and $z$ components:
\begin{subequations}
\begin{equation}
	\tilde{\mathbfss{T}} =
	\begin{pmatrix}
		(\tilde{y}^2+\tilde{z}^2)^{-1} 	& 0 & 0 \\
		0 & (1-\tilde{z}^2)/\tilde{x}^2 & \tilde{y}\tilde{z}/\tilde{x}^2 \\
		0 & \tilde{y}\tilde{z}/\tilde{x}^2 & (1-\tilde{y}^2)/\tilde{x}^2
	\end{pmatrix}
\end{equation}
such that
\begin{equation}
	\tilde{\mathbfss{T}}\cdot\mathbfss{T} =
	\begin{pmatrix}
		1 & -\tilde{x}\tilde{y}/(\tilde{y}^2+\tilde{z}^2) &
			-\tilde{x}\tilde{z}/(\tilde{y}^2+\tilde{z}^2) \\
		-\tilde{y}/\tilde{x} & 1 & 0 \\
		-\tilde{z}/\tilde{x} & 0 & 1
	\end{pmatrix}
\end{equation}
\changed{is identical to unity in this 2x2 sub-matrix.} This gives
\begin{align}
	W' &= W -\frac{\tilde{z}}{\tilde{x}}V_{{\rm p},x} = W - \beta_{U,\mathrm{inv}}V_{{\rm p},x} \label{equ:inv} \\
	L_z' &= L_z
		+\frac{\tilde{x}\tilde{y}r}{\tilde{y}^2+\tilde{z}^2}(\tilde{y}V_{{\rm p},y}+\tilde{z}V_{{\rm p},z})
		-\frac{\tilde{y}X}{\tilde{x}}V_{{\rm p},x}
\end{align}
\end{subequations}
with similar error properties to the previous approach. In particular the error in $L_z$ is random, but does not vanish in the limit $(\tilde{y},\tilde{z})\to0$ of the centre or anti-centre directions.

\subsubsection{Method comparison}
Fig.~\ref{fig:coeff} analyses the coefficients $\beta_i$ for the bias in $W'$ given in equations (\ref{equ:diag}) (two values) and (\ref{equ:inv}) (1 value), when we dissect the sample in $\Lz$ calculated via the basic method.

The top panel of Fig.~\ref{fig:coeff} shows the mean values of $\beta_i$ in each bin. Due to the good sample symmetry, the coefficients are quite small, but explain the subtle differences between the basic and the frame corrected method in Figs.~\ref{fig:meanstat} and \ref{fig:TgasVW}. Since $\beta_{U} \approx -0.02$ across the sample, accounting for the radial reflex motion of the Sun explains the small $\approx 0.2 \kms$ difference we found in the overall vertical motion between the two samples. Further, the larger $\beta_{U}$ near $\Lz \approx 1700 \kms$ exacerbates the dip in $\meanW$ for the basic method by of order $0.2 \kms$, explaining the mild difference. Any bias from in-plane streams like Hercules (offset by $\approx 30 \kms$ for a small sub-group of stars) will be of roughly the same size or smaller.

The bottom panel of Fig.~\ref{fig:coeff} controls for a potential bias by Galactic rotation correlating with the $\beta_i$ coefficients. For this purpose, we multiply the $\beta_{U}$ coefficients with $\sin{\alpha}$, where $\alpha = \tan^{-1}(Y/X)$ is the in-plane angle between the Sun-Galactic Centre and star-Galactic Centre lines, and the $\beta_{V}$ coefficient with $(1-\cos\alpha)$. As we can see, the resulting values are so small that even a multiplication with $300 \kms$ will not affect our results.

\begin{figure}
\epsfig{file=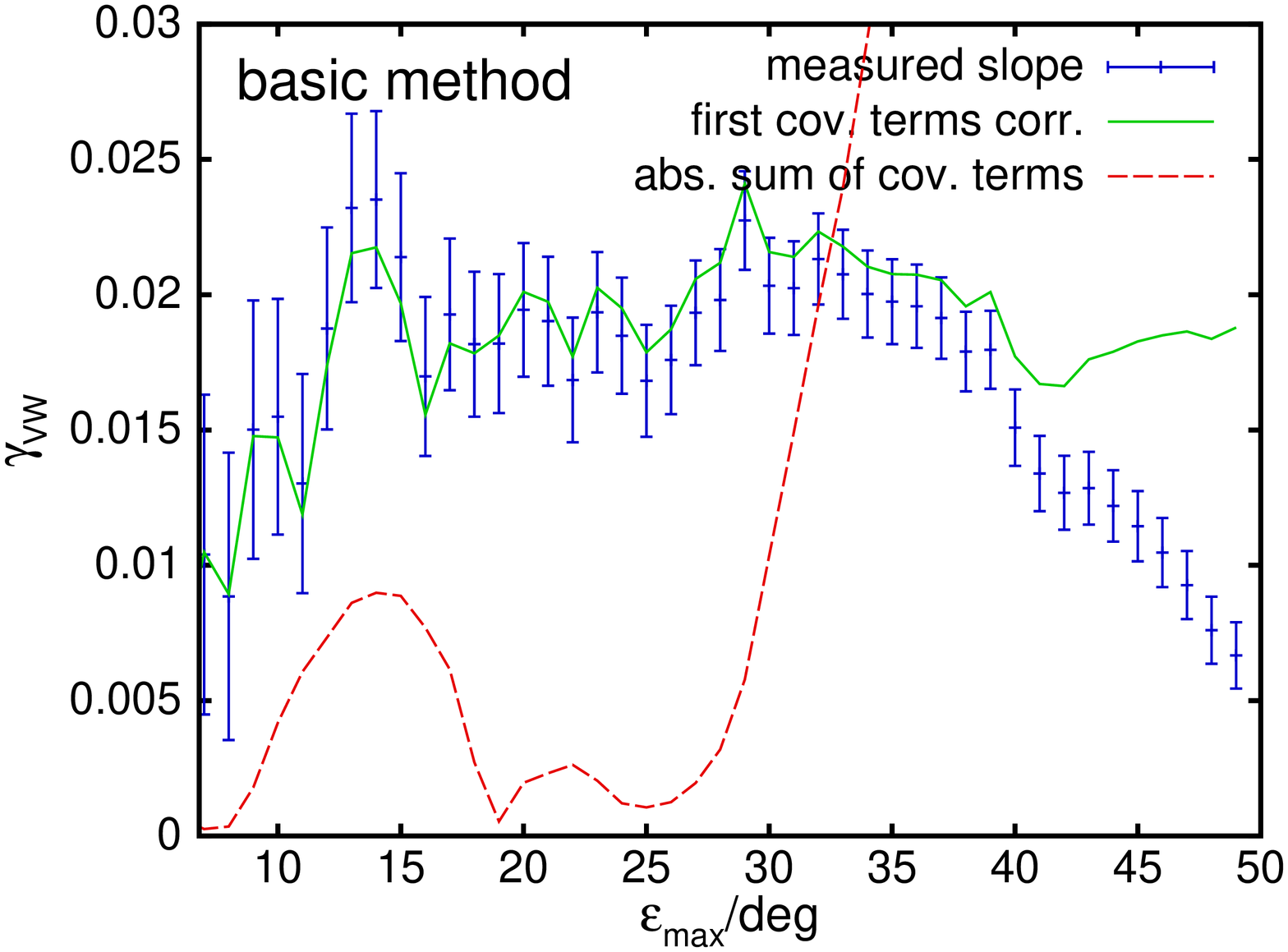,angle=-90,width=\hsize}
\epsfig{file=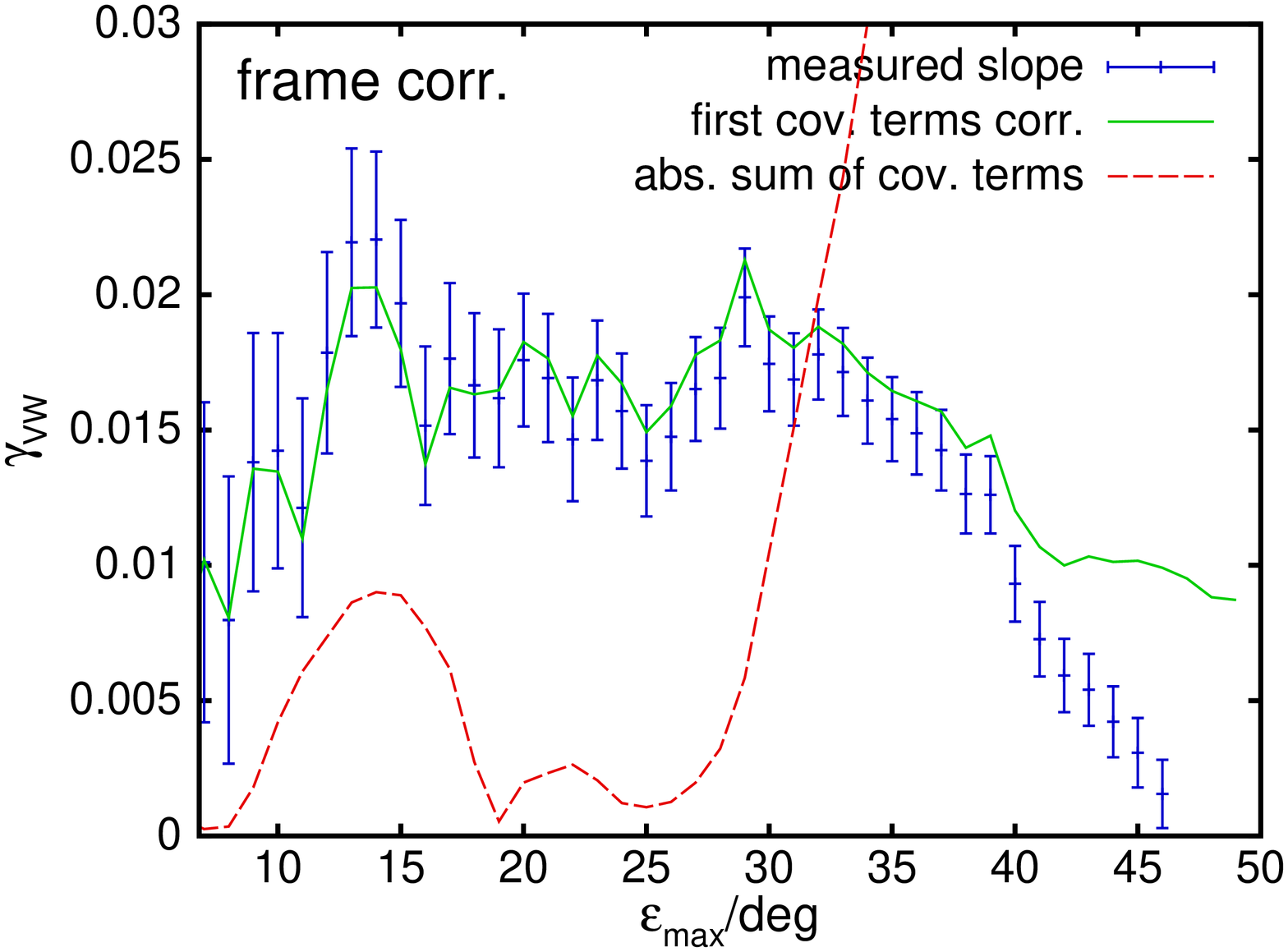,angle=-90,width=\hsize}
\caption{Slope $\gammavw$ of the mean vertical velocity $\meanW$ vs.\ $\Vphi$ obtained from the full TGAS sample for different acceptance angle $\epsilon_{\max}$ ($x$-axis) for the basic method (top) and the frame correction method (bottom). Error bars represent the blank statistics (no trend correction), while the green curve shows the same trend corrected for the leading terms (covariances involving the same velocity component on both sides) from equation~\eqref{eq:fincov}. The red dashed line shows the summed absolute values of these terms.}\label{fig:gradstat} 
\end{figure}

\section{Measuring the Warp}
We use equation~\eqref{eq:fincov} to assess the significance of the trend between $W$ and $\Vphi$ on the full TGAS sample. In Fig.~\ref{fig:gradstat} we show for the basic and the frame correction method, the dimensionless slope of a linear regression of $\overline{W}$ vs. $\Vphi$, $\gammavw = \Cov(W,\Vphi) / \Cov(\Vphi,\Vphi)$, obtained for different acceptance angles $\epsilon$ in galactic coordinates: only stars with $|\gb|<\epsilon$ and $|\bar{\gl}|<\epsilon$ are selected. As detailed above, the covariances $\Cov(W,\Vphi)$ and $\Cov(\Vphi,\Vphi)$ cannot be directly measured. The blue error bars show the favoured estimate via $\Cov(V'',W'')$. The green line shows the same value after correcting for the dominant terms in equation~\eqref{eq:fincov}, which have the same velocity component on both sides. As the mean velocities are comparably small, we can approximate them as $\Cov(aU,bU) \sim \Cov(a,b) \sigma^2_U$, using for this plot $(\sigma_U, \sigma_V, \sigma_W) = (40, 35, 20) \kms$, and measuring the necessary covariances $\Cov(\Tij,\Tkl)$ directly from the data. These corrections partly cancel each other, so we show in addition the sum of their absolute values with a dashed red line. 

As explained in Section~\ref{sec:theory:method}, the polluting terms in equation~\eqref{eq:fincov} cancel out to lowest order if the sampling is symmetric in $\gl$ and $\gb$ around the centre-anticentre directions. However, we expect minor deviations from perfect symmetry in the Gaia-TGAS dataset due to the inhomogeneous scanning law, or due to patchy reddening and crowding losses. These make smaller selections more vulnerable to asymmetries than larger sizes, as exemplified by the excursion of the correction terms near $\epsilon \sim 14 \degr$. However, for $\epsilon = 30 \degr$ the sample is almost perfectly symmetric. \changed{Consistent with that the difference between the mean $z$ and $y$ coordinates of each subsample and the Sun is less than $0.02 \kpc$.}  

In Fig.~\ref{fig:gradstat} the slope $\gammavw$ of $\meanW$ against $\Vphi$ remains stable up to $\epsilon \sim 30 \degr$. The solid green curve demonstrates that the correction terms capture most of the deviations, both for the excursion near $\epsilon \sim 14 \degr$ and towards large $\epsilon$. Consistent with the stability out to $\epsilon \sim 30 \degr$, these corrections are negligible. Beyond $\epsilon > 35 \degr$ the approximation breaks down and the measured slope starts to deviate from its well-determined value. This is correctly signalled by a rapid increase in the first order correction terms, which hold the slope stable out to $\epsilon \sim 40 \degr$. The correction terms were calculated for the basic derivation and hence work better than for the frame corrected method (bottom). We also note that the basic method shows better stability to large $\epsilon$. 

\begin{figure}
\epsfig{file=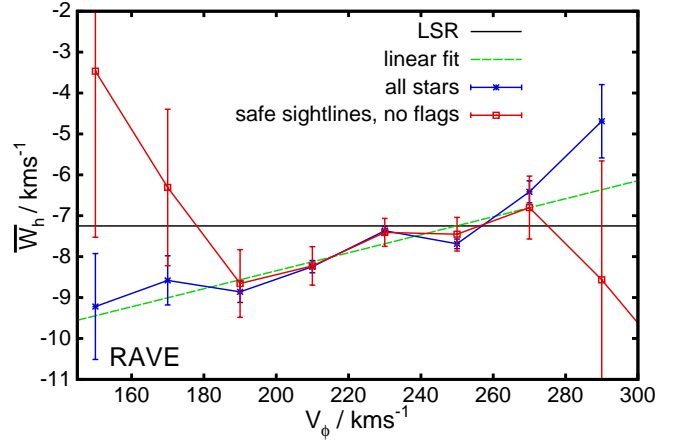,angle=-90,width=\hsize}
\caption{Mean heliocentric $W$ velocity in the RAVE-TGAS sample binned by $\Vphi$. }\label{fig:RaveVW}
\end{figure}

While dealing with a smaller number of stars, the danger of systematic biases for the RAVE sample is more benign: systematic cross-terms between the velocity components would only arise under systematic distance errors. However, SA17 have shown that the distances we employ here are free of significant systematic biases. This is confirmed by Fig.~\ref{fig:RaveVW}, where we plot $\meanW$, averaged over bins in $\Vphi$, for all RAVE stars passing our quality cuts with blue error bars, and all stars within ``safe sightlines'' (selecting only stars with small $|\Tvw| < 0.15$) with red error bars. SA17 have also shown that the RAVE $\vlos$ are basically free of systematic or unexpected random errors. The green line shows a linear fit to the full sample. No significant change is detected for the stars in ``safe'' sightlines. The differences at extreme velocities are still consistent with random fluctuations.

To summarize: Fig.~\ref{fig:gradstat} and the stability of mean motions vs. epsilon (see Appendix) demonstrates that an opening angle near $\epsilon \sim 30 \degr$ is optimal for our purpose. More importantly, we have measured the slope in $\meanW$ vs. $\Vphi$, i.e.\ the signal of the Galactic warp, to be $\sim 0.02$ at better than $\sim10\sigma$ formal significance, with a comparable systematic uncertainty.

\begin{figure*}
\epsfig{file=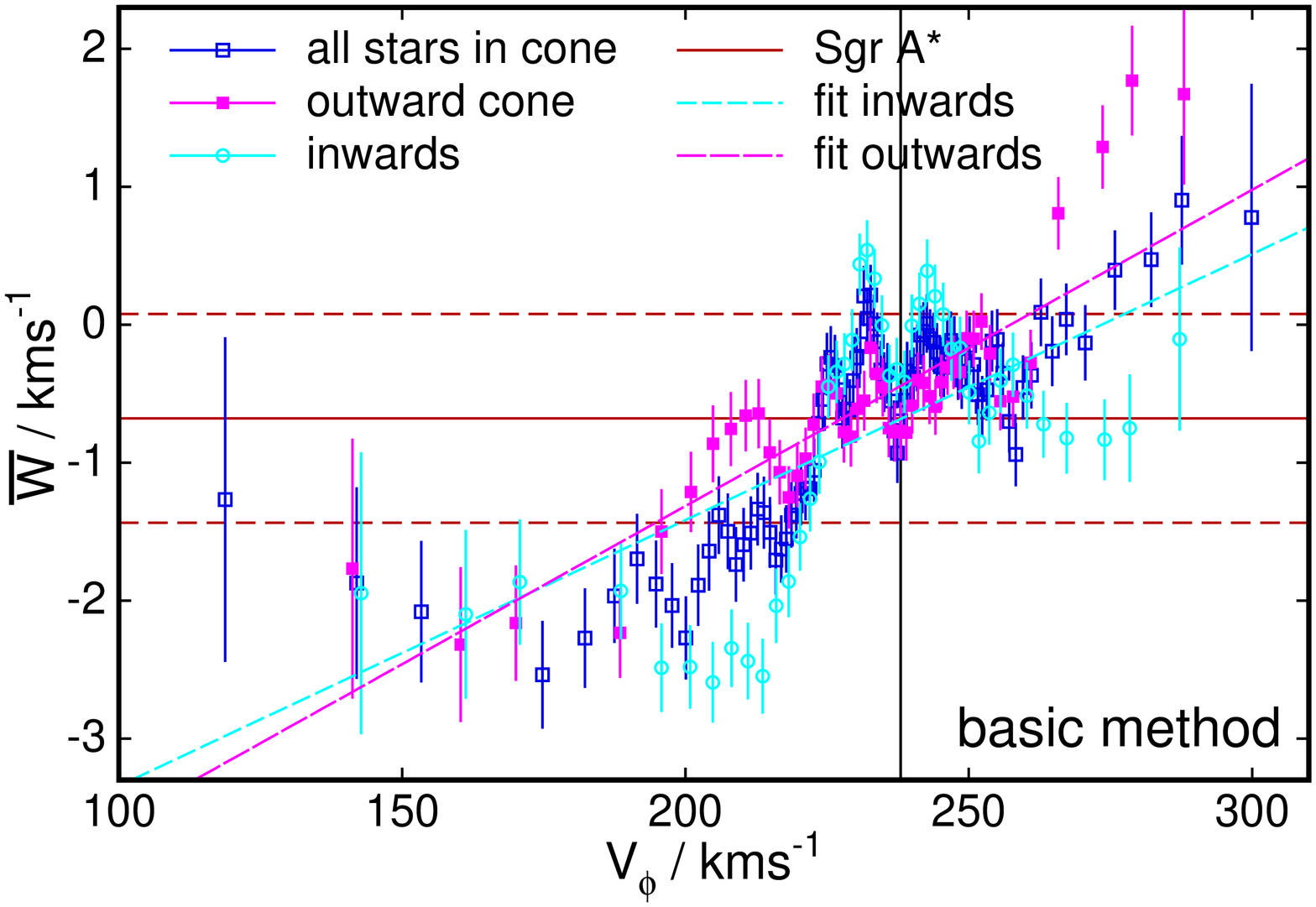,angle=-90,width=0.49\hsize}
\epsfig{file=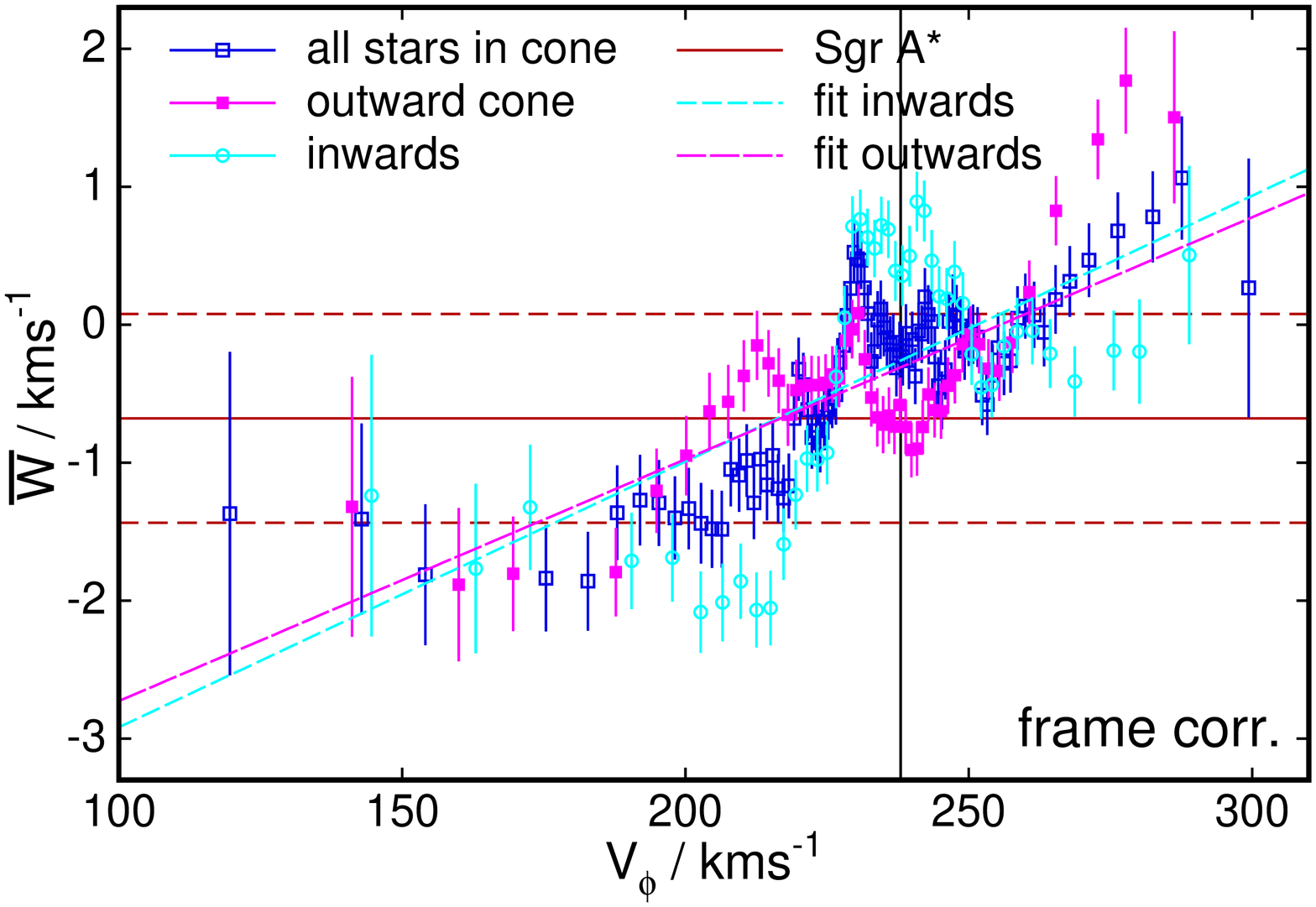,angle=-90,width=0.49\hsize}
\epsfig{file=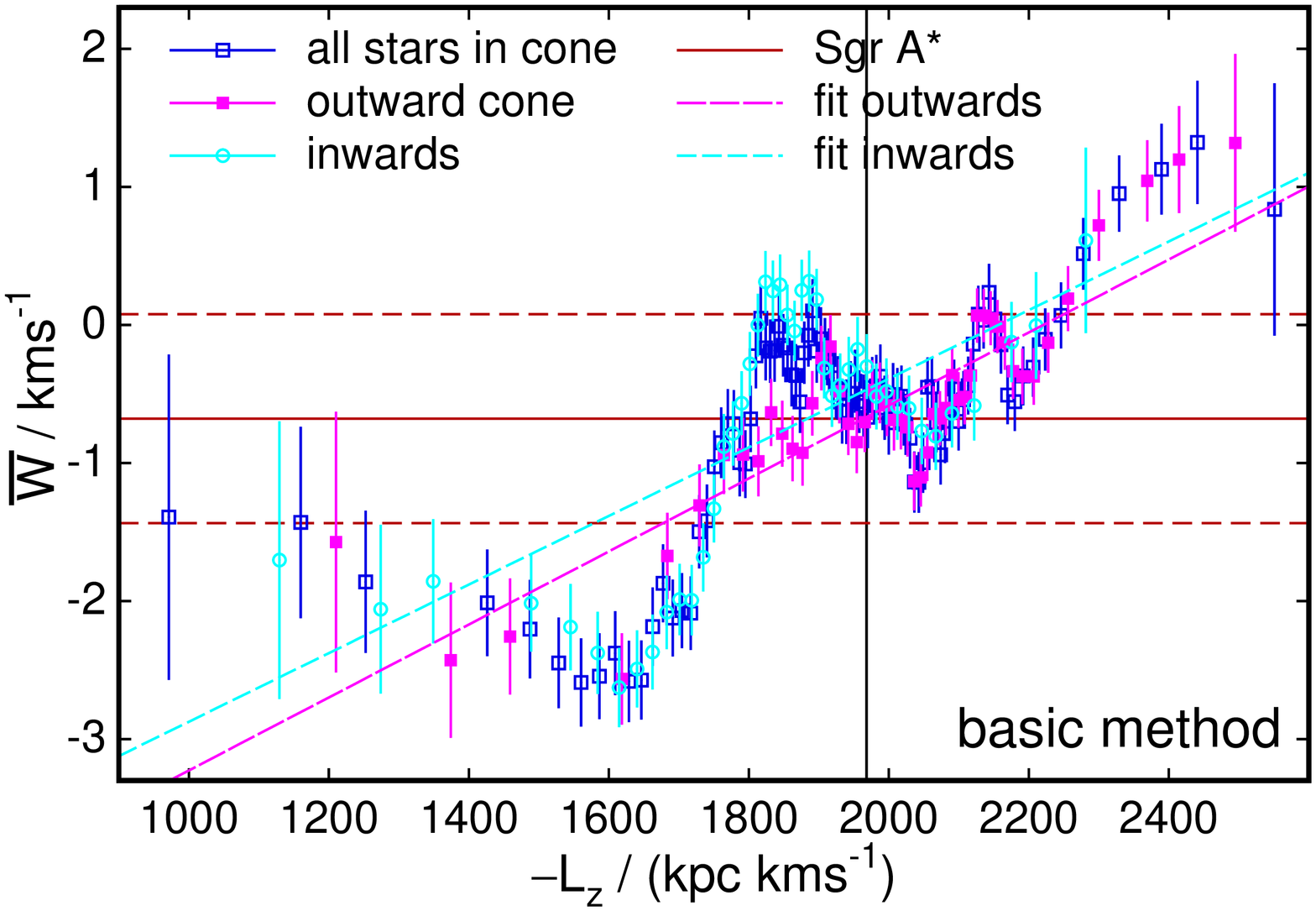,angle=-90,width=0.49\hsize}
\epsfig{file=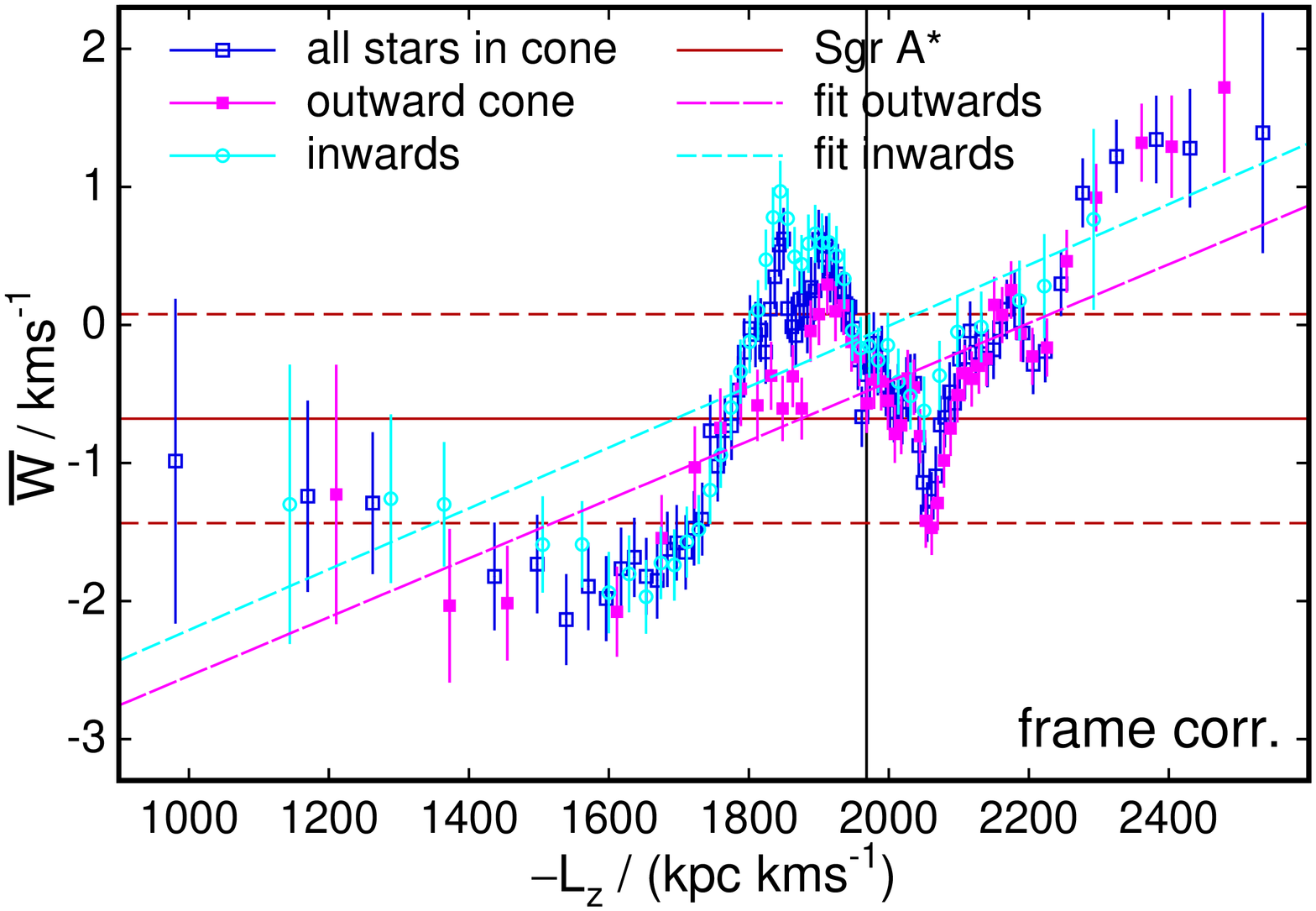,angle=-90,width=0.49\hsize}
\caption{Mean $W$ velocity in the TGAS sample, restricted to an acceptance angle $\epsilon<30\degr$, when binning the sample in azimuthal velocity $\Vphi$ (\emph{top}) or angular momentum $\Lz$ (\emph{bottom}). Uncertainties are estimated from the velocity dispersion measured in each bin. Only every third data point is independent, since we let a mask of width $6000$ slide over the sample in steps of $2000$. Each panel shows the statistics for the full sample (blue) or samples restricted to the outward and inward cones \changed{(i.e. towards the anti-centre and centre directions)}. Sloping lines are linear fits at $\Vphi > 100 \kms$ and $-\Lz > 800 \kpc \kms$, respectively. Horizontal lines indicate the vertical velocity frame defined by Sgr\,A$^\star$ and its uncertainty, the vertical lines mark the Local Standard of Rest.}\label{fig:TgasVW}
\end{figure*}

\begin{figure}
\epsfig{file=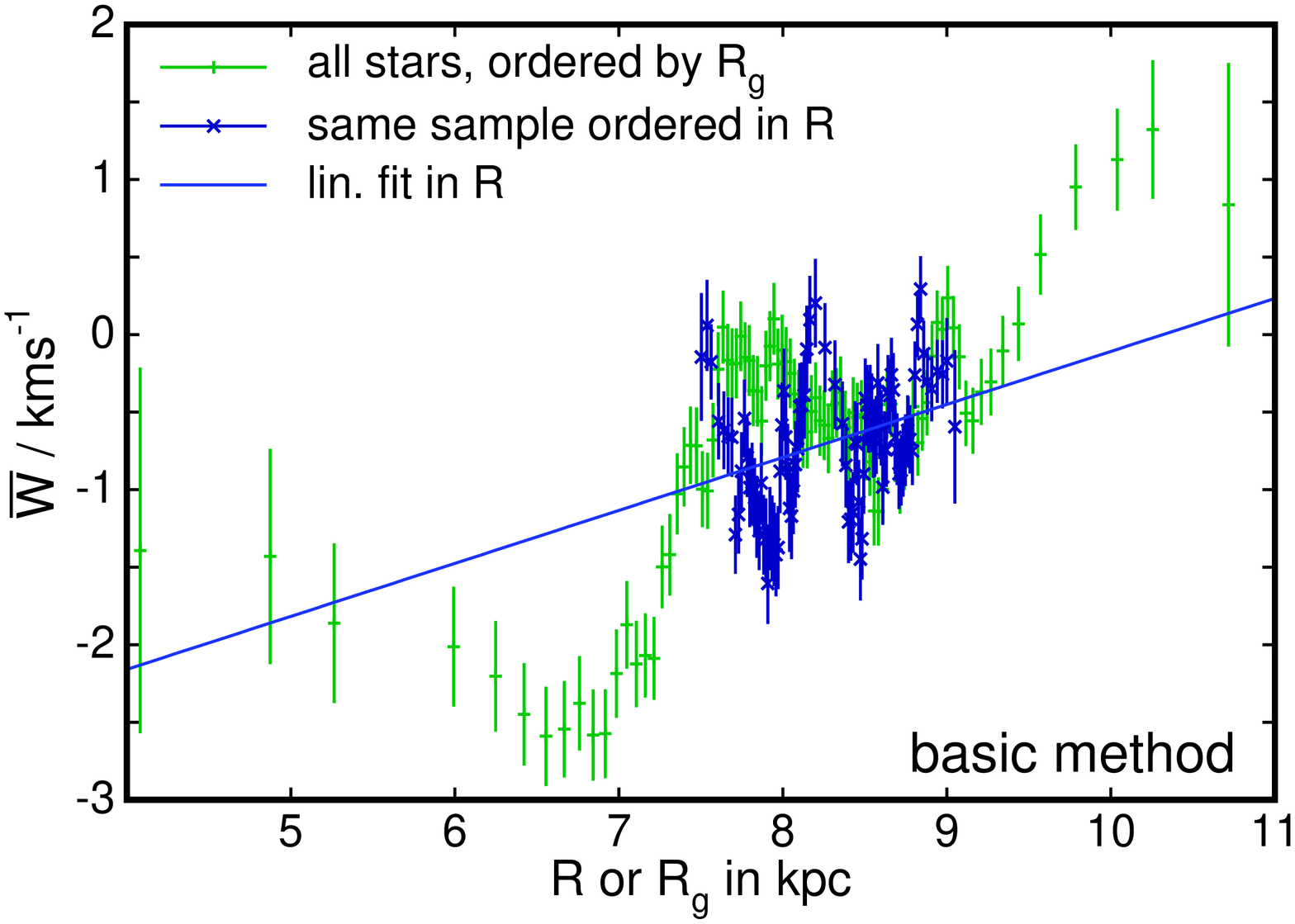,angle=-90,width=\hsize}
\caption{We show the direct correlation between mean $W$ motion and Galactocentric radius $R$ of the stars, shown for the basic method. Just like in Fig.~\ref{fig:TgasVW}, we order the sample in the $R$ or respectively $\Rg$ and \changed{then use overlapping bins of 6000 stars in steps of 2000} across the sample. The green error bars show the guiding centre radius $\Rg$, whereas the dark blue error error bars display the sample separated in Galactocentric radius $R$. The line depicts a linear fit to the latter and has a slope $(0.34 \pm 0.12) \kms/ \kpc$.} \label{fig:Rwmap}
\end{figure}

\section{Detailed Kinematics}\label{sec:measure}
Fig.~\ref{fig:TgasVW} shows the mean vertical velocity of stars in the TGAS sample vs. azimuthal velocity $\Vphi$ (top) and angular momentum $\Lz$ (bottom) respectively. For this plot we sort the sample in either $\Vphi$ or $\Lz$ and then let a mask of width $6000$ slide over the sample in steps of $2000$, so only every third data point is independent and at the extreme ends of the range samples contain only $\sim 4000$ and $\sim 2000$ stars. The data points show, for each bin, the mean vertical velocity vs.\ the mean value of $\Vphi$ or $\Lz$ along the abscissa. At the lowest $\Vphi$ values, bins also contain halo stars; however, their number is small in TGAS. The error bars are set by $\sigma_W/\sqrt{N}$, where $\sigma_W$ is the measured dispersion in the vertical component and $N$ the number of stars in each bin. In each plot we show the statistics for the full sample (dark blue), as well as splitting it into the outward and inward cone. The sloping lines show linear fits to each the inward and outward cone. In all samples, we find a highly significant slope of the vertical velocity vs. both $\Vphi$ and $\Lz$. In light of these findings we cannot fully trust the vertical Local Standard of Rest determination from local stars. Hence, we plot with horizontal lines the Galactic Standard of Rest that would be implied if we assume Sgr A$^\star$ to be at rest, using the proper motion from \cite{Reid04} and the Galactocentric distance from \cite{S12}. The errors are $1 \sigma$ errors assuming full independence of the proper motion and distance measurements. The motion of Sgr A$^\star$ actually indicates that the Local Standard of Rest \changed{as defined by stars in the Solar Neighbourhood} is moving upwards by $\sim 1 \kms$, i.e. the Solar position is still affected by the warp signal. This is consistent with the fact that the decrease of $W$ toward smaller $\Vphi$ (top panel of Fig.~\ref{fig:TgasVW}) continues all the way to $\Vphi\approx200\kms$, which is dominated by stars with guiding centres within the Solar circle.

If the trends in Fig.~\ref{fig:TgasVW} are linked to a galactic warp with a linear $\meanW$-$L_z$ relation, we expect $\Vphi$ to shear by about $20\kms$ in $W$ vs. $\Vphi$, since the outward cone is at $\sim 10\%$ larger average Galactocentric radius $R$ than the inward cone. And indeed we observe a clear hint that the slope of $W$ vs. $\Vphi$ starts at $\Vphi \sim 215 \kms$ in the inward cone, but at significantly lower $\Vphi$ in the outward cone. Similarly, the mean $W$ velocities at high $\Vphi$ are significantly larger in the outward cone. 

Remarkably, the differences between the inward and the outward cones largely disappear when plotting $W$ against angular momentum (bottom panel of Fig.~\ref{fig:TgasVW}), consistent with interpretation that we are dealing with large-scale structure, and not some influence of local streams.

Both panels show a local peak in the mean $W$ motion near the Solar $-\Lz\approx 2150 \kpc\,\kms$. It is not yet clear if this is a statistical fluctuation, caused by a local stream, a potential issue with proper motions, or a physical feature of the warp. Inspection of the detailed $\Vphi$-$W$ velocity planes in Fig.~\ref{fig:vwmap} does not reveal any obvious streams. Similarly, when we plot the residuals from a Gaussian fit in Fig.~\ref{fig:sgauss}, we still do not see any structures appear (apart from the fact that a Gaussian is poor fit to velocity distributions). On the other hand, there is no firm expectation for the warp signal to be smooth in $\Vphi$. This question will have to await better statistics from Gaia DR2.

Having established the firm trend of $\meanW$ with $\Lz$, one may of course ask if the same effect cannot be detected in Galactocentric radius $R$ directly. This comparison is shown in Fig.~\ref{fig:Rwmap}, where we plot the mean vertical motion $W$ vs. Galactocentric radius $R$ on the abscissa for the ``basic'' method. Similar to Fig.~\ref{fig:TgasVW}, we order the samples in either $R$ or $\Rg$ respectively, and then let a selection bin of width $6000$ move in steps of $2000$ each over the sample. Due to the small extent of the Gaia DR1 sample, the short baseline in $R$ makes it hard to reliably detect any signal. Also, the region covers mostly the downward slope of the short-wave component in $\Rg$, so no firm trend is expected. Formally, an uptrend in radius is detected using the ``basic'' and ``inversion'' methods with about $0.3 \kms\kpc^{-1}$, but no significant result is obtained in the frame corrected data. The results are consistent with the dependence of $\meanW$ on $\Lz$. One should not over-interpret the detection of this weak trend of less than $3 \sigma$. In particular, dissecting the sample directly in $R$ breaks the symmetry of $\cos \gl$. Also, the different slopes should differ to some extent, because at each $R$ the local stellar populations originate from a wide distribution of guiding centre radii. Nevertheless, it offers some additional indication for large-scale structure, and will be of interest in Gaia DR2.\footnote{Our results, e.g.\ in Fig.~\ref{fig:Rwmap}, are quite different from those presented in Fig.~16 of \cite{Poggio17}, who analyse 758 OB stars from the Hipparcos catalogue \citep{HIP2} with Gaia DR1 astrometry. While the 
problem with distances is shortly discussed in their text, the figure in their paper does not account for distance biases or quality cuts. In particular, large distance overestimates (of $\sim100\%$) should dominate their result at $|R-R_0|\gtrsim1\,\kpc$, which explains the large negative $\meanW$ found in their study.}

\begin{figure}
\epsfig{file=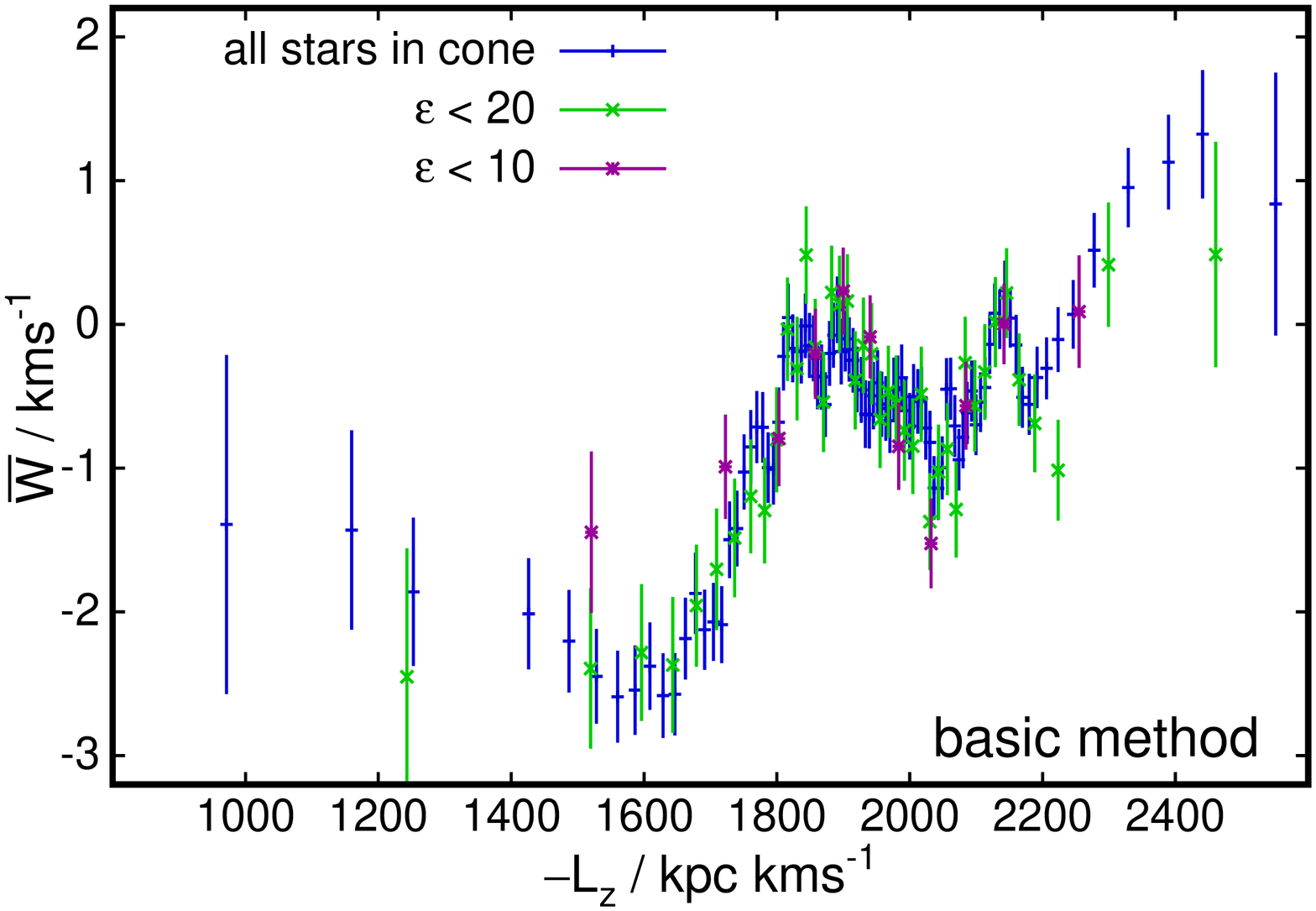,angle=-90,width=\hsize}
\epsfig{file=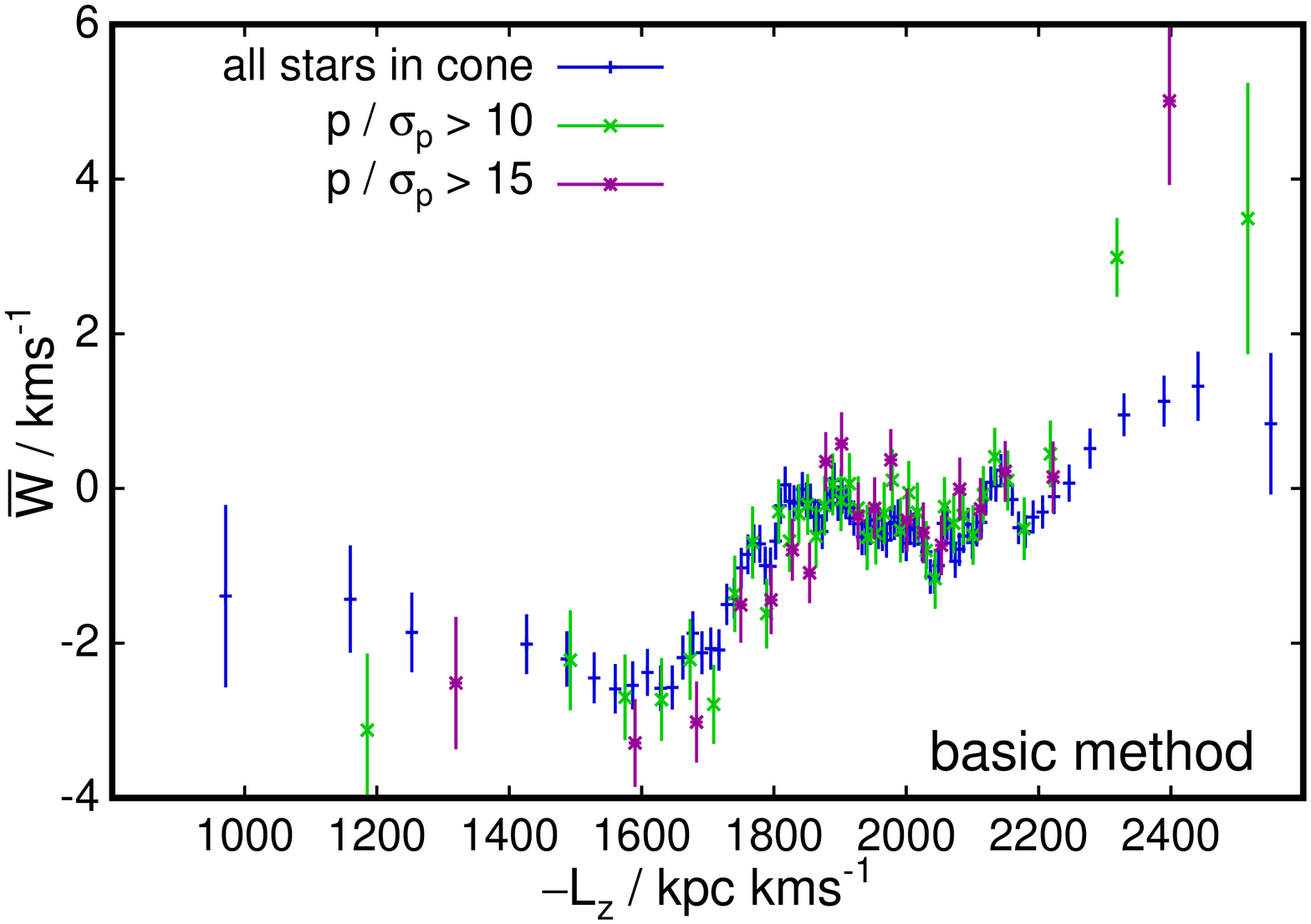,angle=-90,width=\hsize}
\caption{We test the effects of varying the acceptance angle $\epsilon$ for the observational cone (\emph{top}) at values $\epsilon = 30\degr,\,20\degr,\,10\degr$, and of varying the parallax quality cut $p_0/\sigma_p = 5,\,10,\,15$ (\emph{bottom}), shown for the basic method. In each case we show the full sample with the usual sliding $6000$ stars wide bins, but the restricted samples with a $2000$ stars per bin, still moving in steps of $2000$.}\label{fig:testingepsp}
\end{figure}

\begin{figure}
\epsfig{file=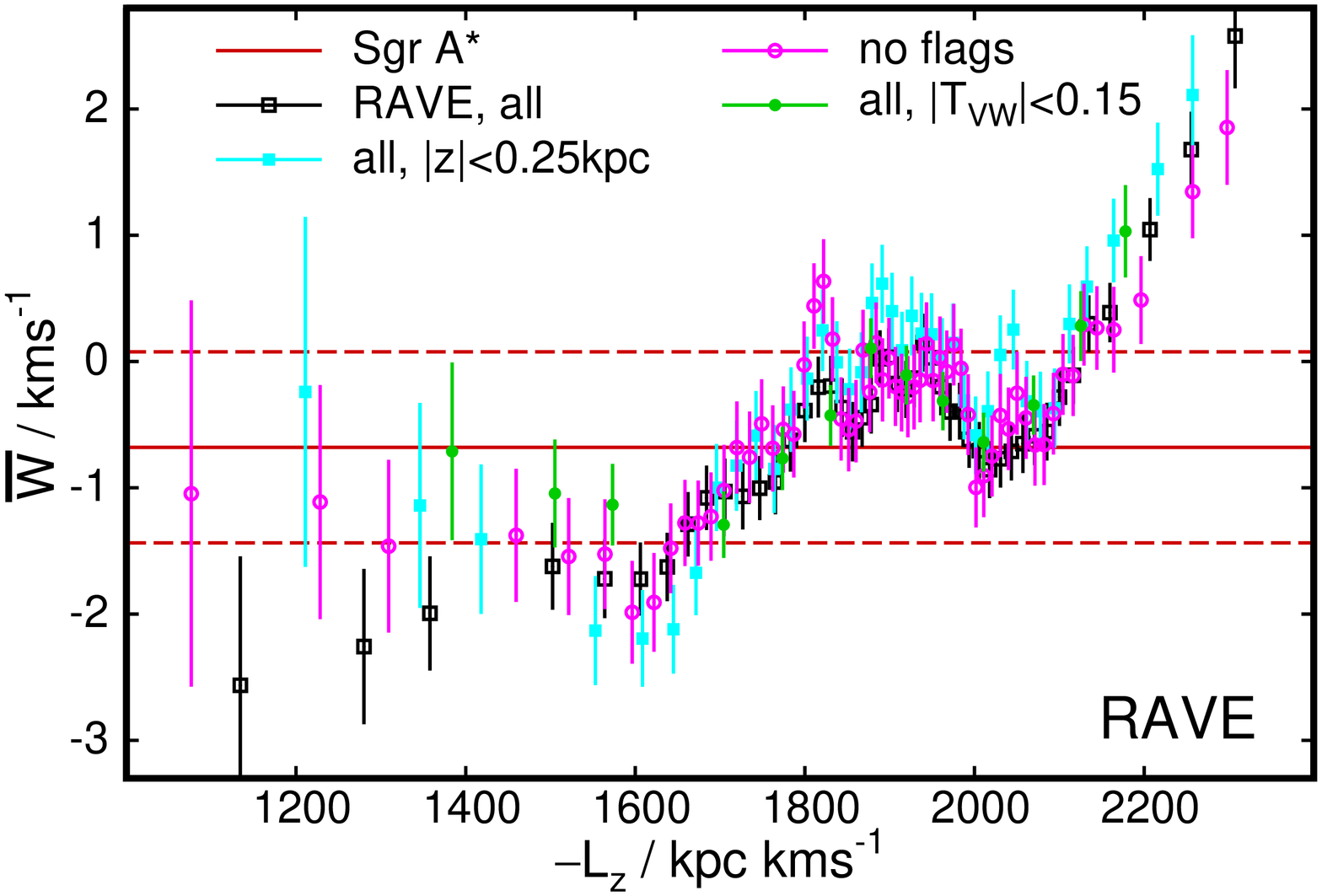,angle=-90,width=\hsize}
\caption{$\meanW$ vs. $\Lz$ for RAVE stars. To resolve any structures in the smaller sample size, we have reduced the bin size to $2000$ stars. Each bin is independent. The different colours indicate sample selections for low altitude $|z|$, for unflagged stars, and for ``safe'' sightlines. The observed structure is consistent with Fig.~\ref{fig:TgasVW}.}\label{fig:RAVEW}
\end{figure}

\subsection{Stability tests and comparison with RAVE-TGAS}\label{sec:disterr}

As discussed in Section~\ref{sec:theory}, the most important source of bias is the missing $\vlos$ information in Gaia-TGAS; we can test for resulting biases by varying $\epsilon$. However, random distance errors can have some impact when selecting/binning the sample in $\Vphi$: in our observational cones, the estimates of both $V$ and $W$ relative to the Sun, i.e. $V-\Vsun$ and $W-\Wsun$, are nearly proportional to $s/s_0$, the ratio of measured to true distance. Random distance errors let stars with overestimated distances concentrate in the wings of the estimated $V$ distribution, while stars with estimated $V\sim\Vsun$
more likely have underestimated distances. Since the Sun is moving towards the North Galactic Pole with $\lesssim 8 \kms$, both wings of the $\Vphi$ distribution should have a negative bias in $\meanW$ \changed{(due to the prevalence of distance overestimates)}, whereas the centre of the distribution should have a positive bias in $\meanW$. Using mock sample tests, described in Appendix \ref{app:mock}, we estimated the size of this effect to be only $\sim 0.1 \kms$ in both tails, much less than the signal we observe. Nonetheless, we perform further checks by 
varying the cut of $p_0/\sigma_p$.

Fig.~\ref{fig:testingepsp} compares the $\meanW$-$\Lz$ relationship to smaller subsets with tightened cuts either in the acceptance angle $\epsilon$ (top panel) or in parallax quality $p_0/\sigma_p$ (bottom panel). As expected, variation of $\epsilon$ does not yield any significant effect. We further tested that the shape of the selection cones does not matter. Our simple cut in $\gl$ and $\gb$ keeps the symmetry conditions for $\mathbfss{T}$. As expected, selecting a truly circular cone of width $\epsilon$ did not significantly alter our results.

When we increase the minimum parallax quality to $p_0/\sigma_p > 10$, most of the relationship remains identical. However, at very large $|\Lz|$, there is a strong increase in $\meanW$. At first glance this appears consistent with a removal of distance overestimates, which could suppress the increase in $\meanW$ towards larger $\Vphi$. On second glance, however, this is a shaky argument: since the Sun has $W=7.24 \kms$, if the real $\meanW$ of those stars was $\sim 5 \kms$, the average distance overestimate required to lower $\meanW$ to $\sim 2 \kms$ for the $\sim 8000$ stars at the highest $|\Lz|$ is $\geq100\%$, more than an order of magnitude larger than expected from our mock sample tests. The feature is similar to the structure seen in the Hipparcos sample \citep[e.g.][]{Dehnen98}. This may be just a statistical quirk, or it could be caused by the kinematic selection bias in Hipparcos surfacing because Hipparcos stars have better astrometric solutions than the Tycho sample. Real or not, the deviation would physically make sense. We cannot assume that stars at all locations follow a simple tilted ring model pertaining to their guiding centres. Near their peri- and apocentre, they move directly through the gravitational potential of components with a different tilt, and thus \changed{should deviate from the simple picture, in particular in resonant cases} when one of the frequencies of the orbit is similar to the frequency of the perturbation by the warp.

Fig.~\ref{fig:RAVEW} shows a comparison of the $\meanW$-$\Lz$ correlation for different sub-samples in RAVE. The RAVE sample has very different systematics, but is significantly smaller (only up to $60000$ stars pass the quality criteria, compared with $180000$ objects in the $\epsilon < 30 \degr$ cone in Gaia DR1). The survey is asymmetric in $\gb$, pointing predominantly southward, and due to the comparably small number of stars we cannot select stars in a small cone. This results in a potential contamination of our signal with disc breathing modes (which are symmetric in $z$). Moreover, the asymmetry increases the risk of cross-contamination between velocity components, if there is a residual distance bias \citep[][]{SBA, Carrillo18}. On the other hand, RAVE $\vlos$ are excellent, and our distances are statistically very well tested. Fig.~\ref{fig:RAVEW} shows different selections on RAVE: the subsample of unflagged stars (purple, excluding peculiar stars and identified binaries), a low-altitude sample (light blue, $|z| < 0.25 \kpc$), and the very small sample with $|\Tvw| < 0.15$ (green), which limits cross-contamination of $V$ and $W$ velocity components. Each subsample of RAVE-TGAS shows consistently the same pattern as the Gaia-TGAS sample. RAVE points inwards in the Galactic disc, so does not have good coverage at large $|\Lz|$, but it hints for mildly different structure near $-\Lz = 2150 \kpc \kms$ region, where Gaia-TGAS shows one or two narrow spikes in $\meanW$, adding to the suspicion that this feature is a local stream. 

In short: The trend and wave-like pattern found in $\meanW$ vs. $\Lz$ proves robust against the opening angle $\epsilon$ and the parallax error, apart from a deviation at very high $\Lz$, which we interpret as a local deviation, consistent with the steeper warp-related trend observed previously on Hipparcos. The RAVE-TGAS sample is an important comparison due to its very different suspected biases, and confirms the trend and wave-like pattern on Gaia-TGAS.

\begin{figure}
\epsfig{file=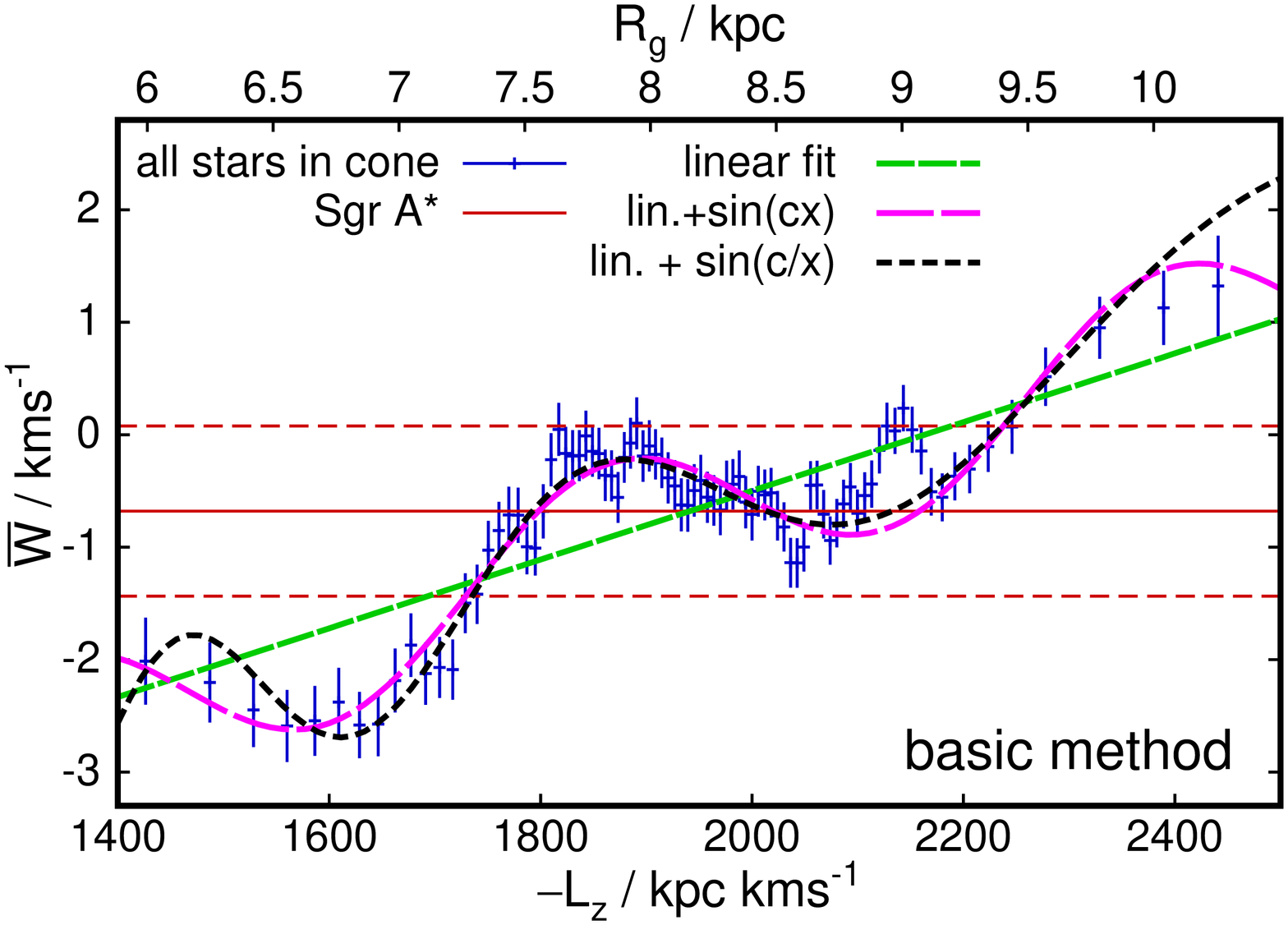,angle=-90,width=\hsize}

\vspace{3mm}
\epsfig{file=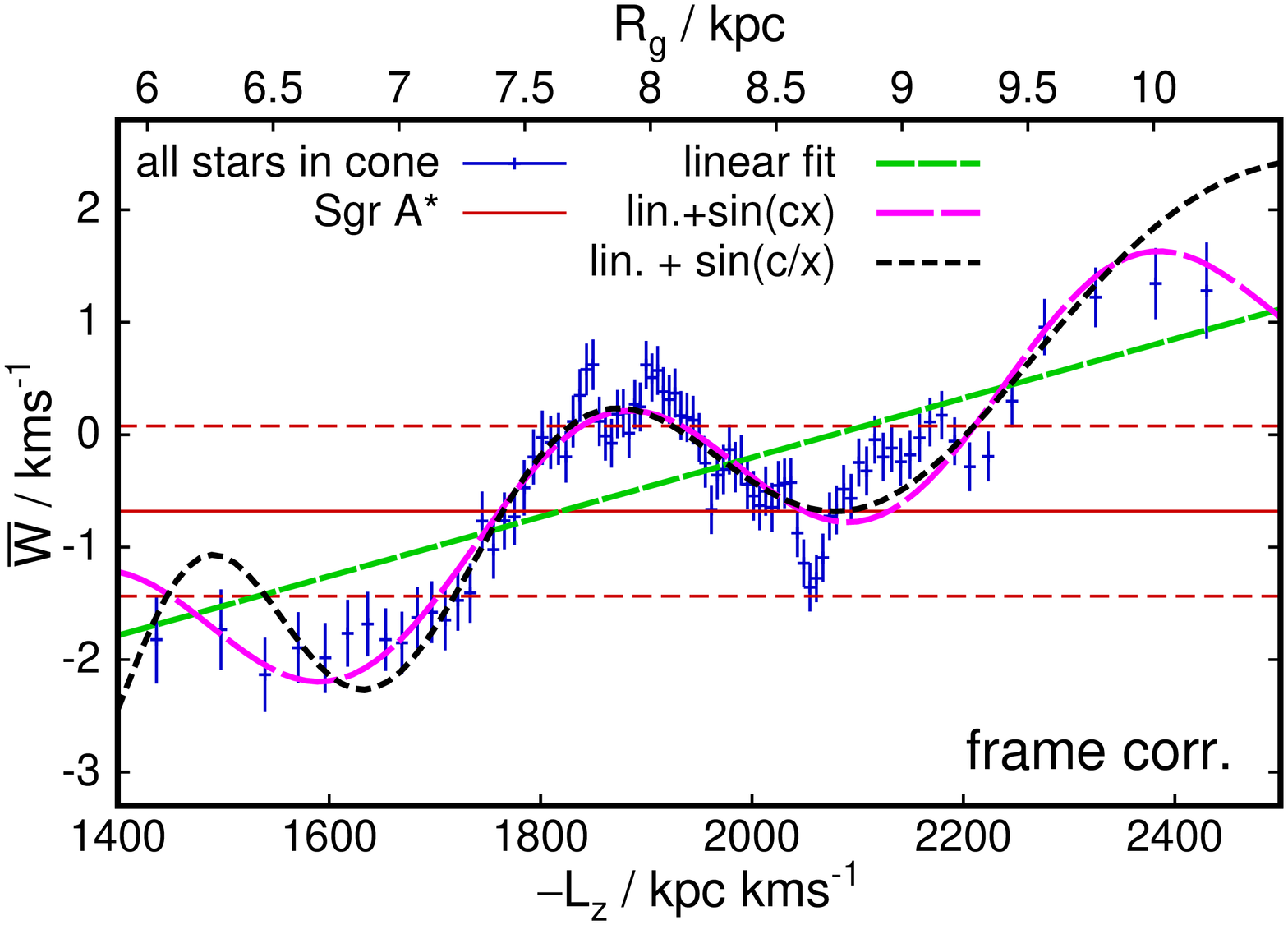,angle=-90,width=\hsize}

\vspace{3mm}
\epsfig{file=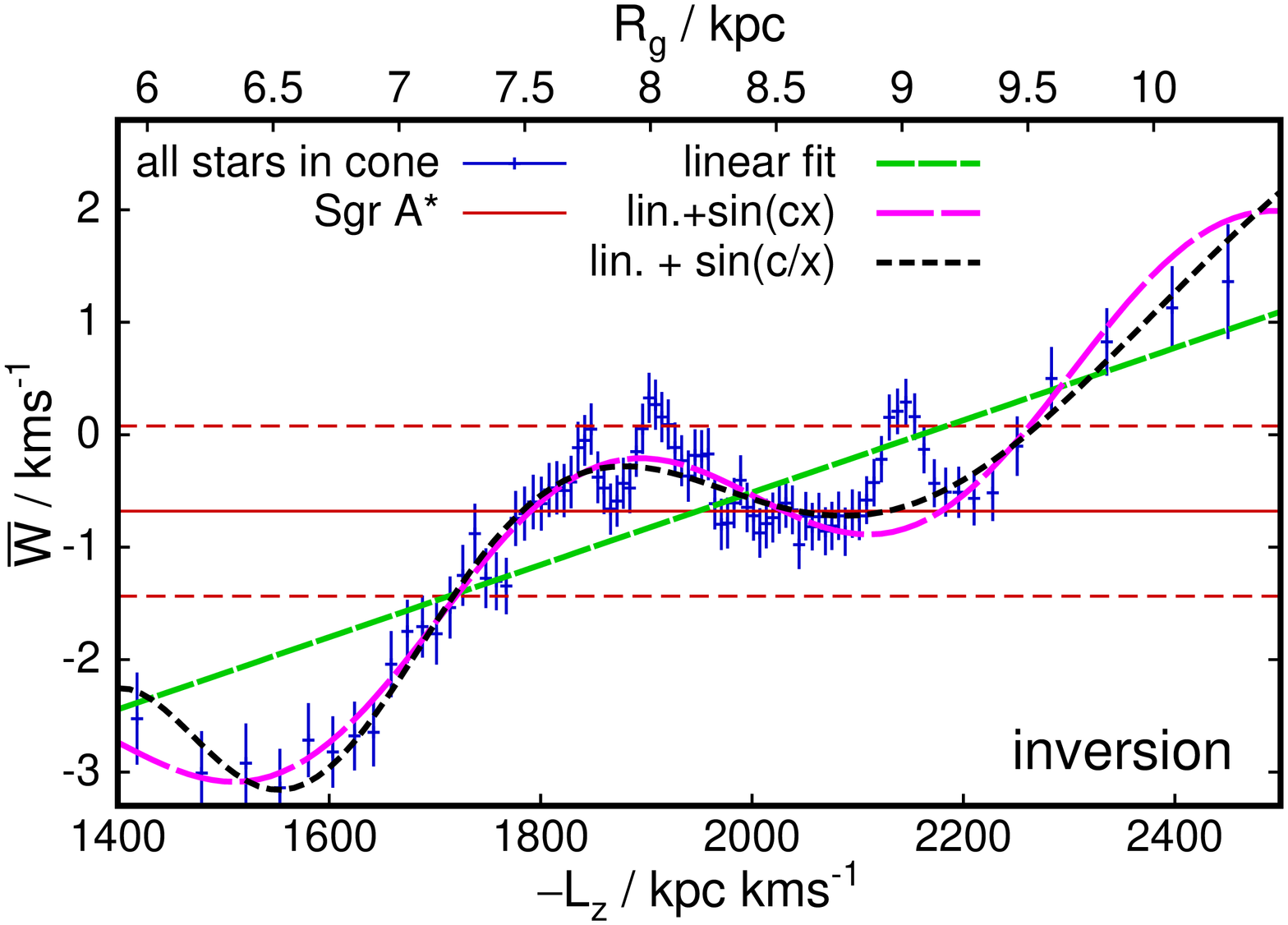,angle=-90,width=\hsize}
\caption{Testing three naive models for describing $\meanW$ vs. $\Lz$ or guiding centre radius $\Rg$ respectively. The lines show linear fits and the fits from equations (\ref{eq:fixsin}) and (\ref{eq:varsin}), the parameters are given in Table~\ref{tab:sinfits}.}\label{fig:sinfits}
\end{figure}

\begin{table*}
\begin{center}
\caption{Best-fit parameters of the functions~(\ref{eq:warpsimple}-\ref{eq:varsin}) fit to the data in Fig.~\ref{fig:sinfits}. $a$ is the slope of the linear trend, $b$ is the $\meanW$ value at $-\Lz = 1600 \kpc \kms$, $c$ sets the length scale of the sinusoidal pattern (for the simple sinusoidal fit, $c$ is the period, and $500 \kpc \kms$ in $\Lz$ translate to $2.1 \kpc$ in $\Rg$), $d$ is the phase, $A$ is the amplitude of the wave-like pattern. All fits are performed for $1400 < -\Lz/(\kpc\kms) < 2400$, excluding a $60 \kpc\kms$ wide region around $-\Lz = 2140 \kpc\kms$.}
\begin{tabular}{l|ccccc}
fit & $a\times 10^{3}/kpc$ & $b/\kms$ & $c/\kpc\kms$ & $d$ & $A/\kms$ \\
\hline \hline
basic &&&&&\\
linear & $3.05 \pm 0.25$ & $-1.721 \pm 0.091$ & --- & --- & --- \\
simple sin. & $3.28 \pm 0.26$ & $-1.846 \pm 0.094 $ & $526 \pm 21$ & $-1.56 \pm 0.18$ & $0.719 \pm 0.070$ \\
wrapping & $3.92 \pm 0.27$ & $-2.011 \pm 0.097$ & $6955 \pm 259$ & $3.03 \pm 0.86$ & $0.764 \pm 0.072$\\
\hline 
frame corrected &&&&& \\
linear & $2.64 \pm 0.25$ & $-1.267 \pm 0.091$ & --- & --- & --- \\
simple sin. & $2.83 \pm 0.25$ & $-1.381 \pm 0.092$ & $499 \pm 17$ & $-1.74 \pm 0.16$ & $0.817 \pm 0.068$ \\ 
wrapping & $3.46 \pm 0.27$ & $-1.534 \pm 0.097$ & $7433 \pm 229$ & $ 1.37 \pm 0.76$ & $0.848 \pm 0.069$ \\
\hline
inversion &&&&& \\
linear & $3.21 \pm 0.25$ & $-1.800 \pm 0.092$ & --- & --- & --- \\ 
simple sin. &  $3.67 \pm 0.28$ & $-2.024 \pm 0.103$ & $600 \pm 27$ & $-1.08 \pm 0.18$ & $0.811 \pm 0.072$ \\
wrapping & $4.32 \pm 0.28$ & $-2.188 \pm 0.100$ & $5723 \pm 246$ & $0.73 \pm 0.83$ & $0.834 \pm 0.076$\

\end{tabular}\label{tab:sinfits}
\end{center}
\end{table*}

\section{Discussion and Comparison to Simple Models}\label{sec:models}

We can think of three (not mutually exclusive) ways to interpret Fig.~\ref{fig:TgasVW}:
\begin{itemize}
    \item{The classic approximation of the Galactic warp, where the Sun is placed close to the line of nodes, and hence the warp appears as a systematic vertical motion of stars, depending on $\Lz$. This feature appears to be overlaid by one of the following two options:}
    \item{The notion of a wave-like structure in the disc, where vertical waves might be propagating radially.}
    \item{The possibility that the observed signal could be a wrapping up perturbation by a merger, e.g. with the more massive halo of the Sagittarius dwarf galaxy.}
\end{itemize}

A simple warp model assumes that the mean vertical component of stellar motion $\meanW$ follows approximately the large-scale warp in $\Lz$ \citep{Dehnen00}. In this simplistic picture the vertical motion in each subsample is given by:
\begin{equation}\label{eq:warpsimple}
    \meanW \sim A(\Lz) (\Omega(\Lz) - \Omega_p)\cos(\gamma),
\end{equation}
where $A$ is the vertical amplitude of the warp, $\Omega(\Lz)$ is the orbital frequency of a circular orbit with $\Lz$, $\Omega_p$ is the angular frequency at which the warp pattern moves, and $\gamma$ is the difference in galactic azimuth between the line of nodes (i.e. where each tilted annulus intersects the Galactic plane) and the position of the star. \cite{Drimmel01} find that the line of nodes of the warp is almost perfectly at the Solar position. While they raise questions about the technical robustness of their fit, we can quite safely assume that $\cos(\gamma) \sim 1$. For small $\Omega_p$, $\meanW = 1\kms$ corresponds to a vertical amplitude of about $35 \pc$. 

Fig.~\ref{fig:TgasVW} demonstrates that a smooth warp according to equation~\eqref{eq:warpsimple} cannot explain the data. 
However, discs can oscillate vertically with (near) radially propagating waves \citep{Hunter69}. In this case, the radial wavelength should be at best constant, or rather decline towards larger radii, since the vertical restoring force, and thus phase velocity of the wave, decreases. We naively fit this by
\begin{equation}\label{eq:fixsin}
    \meanW(\Lz') = b + a\Lz' + A\sin(\Lz' 2\pi/c + d).
\end{equation}
where we moved the zero point of the abscissa by $-\Lz' = -\Lz - 1600.0 \kpc \kms$ towards the onset of the observable warp for convenience.
The alternative is to think of a warp-like feature created by, say, the impact of the Sagittarius dwarf \citep{Ibata98}, or the Magellanic Clouds onto the Milky Way halo \citep{Weinberg95, Weinberg06}. In this case, we could expect the warp to start wrapping up due to the difference in \changed{(vertical)} oscillation frequencies, which can be expected to decrease towards larger radii. We just test a naive fit with
\begin{equation}\label{eq:varsin}
    \meanW = b + a \Lz' + A \sin(-2\pi c/\Lz  + d).
\end{equation}

We plot the fits to equations~\eqref{eq:fixsin} and \eqref{eq:varsin} in Fig.~\ref{fig:sinfits} for all three approximations. The best-fit parameters to the (unbinned) data are shown in Table~\ref{tab:sinfits}. Neither of the two 
interpretations can be trusted, and the fact that both naive models fit strikingly well demonstrates that the current sample extent is too small to detect radial changes of wavelength. 

All analysis methods agree in their fit parameters (see Table \ref{tab:sinfits}) and the observed patterns. The basic method shows slightly lower $\meanW$ around $-\Lz \sim 1600 \kpc \kms$, which is expected from a slight increase in sample asymmetry and a small contamination with radial velocities (see equation \ref{equ:diag} and Fig.~\ref{fig:coeff} in the Appendix). In all cases both the warp and the vertical oscillation are detected at better than $10\sigma$, in particular all methods agree very well on the amplitude of the oscillation of about $0.8 \kms$. The wavelength of the oscillation is mildly longer in the inversion method, but translates to about $2 - 2.5 \kpc$ in $\Rg$. The phase differences (parameter $d$ in equations~\ref{eq:fixsin} and \ref{eq:varsin}) are caused by the small changes in the wave-length, and are irrelevant to the interpretation.

The wave-length of the observed pattern resembles the simulations by \cite{Onghia16}, though we notice that the simulations by \cite{Vega15} do not show such a clear pattern in the vertical velocity vs. in-plane position. In particular, their waves appear to show near-zero net vertical velocity in the disc plane. We also stress again that the observed pattern is consistent with bending waves and very different from the breathing modes associated with spiral arms \citep[see][]{Weinberg91, Monari16}, for which indications have been found in the RAVE survey \citep{Siebert11, Williams13, Faure14}. Breathing modes are even functions around the disc mid-plane with zero vertical mean motion and are filtered out by the north-south symmetry of our analysis.

One should also note the pronounced spike at $-\Lz \approx 2150 \kpc \kms$, corresponding to a guiding centre radius of $9\kpc$ (assuming a flat rotation curve). It is statistically highly significant in our outward cone in Gaia-TGAS. Our favoured explanation is the presence of a stream with either some minor vertical motion, or contamination from strong radial motion. However, resonant interactions of stellar orbits with a warp are unexplored, and might give rise to similar features. We also note that in the RAVE-TGAS sample the strong rise in $\meanW$ vs.\ $|\Lz|$ sets in at around these values of angular momentum, i.e.\ a bit earlier than for Gaia-TGAS. However, the RAVE comparison could be affected by breathing modes owing to the asymmetry of the survey volume.

\section{Conclusions}\label{sec:Conclusions}

In this paper we have derived distances to TGAS stars in Gaia DR1, which, unlike most previous attempts, are by construction unbiased in the mean estimated distance. We have used these distances in conjunction with the TGAS proper motions to measure the kinematic signal of bending modes (including the warp) of the Milky-Way disc in the mean vertical stellar motions. To derive unbiased distances for Gaia-TGAS stars, we have measured an effective spatial selection function for this sample, which is very well described by a simple declining exponential with scale length $0.2\kpc$. The derived distances are freely available\footnote{The distance dataset is online at: \url%
{http://www-thphys.physics.ox.ac.uk/people/RalphSchoenrich/data/tgasdist/tgasdist.tar.gz}.}.

Since the line of nodes of the Galactic warp is close to the Solar position, it imprints on stellar kinematics as an increase of the mean vertical velocity, $\meanW$, with orbital angular momentum $|\Lz|$ (or azimuthal velocity $\Vphi$ at fixed Galactocentric radius $R$). In a sample lacking line-of-sight velocity information, like Gaia-TGAS, these two velocity components can only be accurately determined in the Galactic centre and anti-centre directions. To allow for the use of sizeable cones, we have developed and analysed a set of first-order correction strategies, which provide an unbiased correlation between $\Vphi$ and $W$ and stabilise $\meanW$ for cone opening angles $\epsilon \lesssim 30 \degr$.

The kinematics obtained in this way show a correlation between $\meanW$ and either $\Vphi$ or $|\Lz|$ at more than $10\sigma$ formal significance, with a mean uptrend $\diff\meanW /\diff|\Lz| = (3.05 \pm 0.25)10^{-3} \kpc^{-1}$ or $\diff\meanW / \diff \Vphi \sim 0.02$ (Fig.~\ref{fig:TgasVW}), and systematic uncertainty of the same order. The RAVE-TGAS sample, which contains line-of-sight velocities, shows a consistent trend (Fig.~\ref{fig:RaveVW}).

In contrast to previous studies, which placed the onset of the warp mostly at or outside the Solar Galacto-centric radius $R_0$, we find that the vertical mean motion associated with the warp commences at $\Lz$ corresponding to guiding centre radii inside the Solar circle at $\Rg \lesssim 7\kpc$, similar to what has been found from near infrared photometry \citep{Drimmel01}.

We find that a smooth, monotonous warp pattern cannot satisfactorily describe the data. Instead, both the inward and outward cone show a significant wave-like pattern on top of the warp signal. Strikingly, while the inward and outward cone do not match in velocity space, the patterns perfectly agree in $\meanW$ vs.\ $\Lz$, demonstrating that the wave-like pattern directly connects $\meanW$ to $\Lz$ of the stars. The pattern is well described by a simple sinusoidal wave with a wavelength of about $2.5 \kpc$ in guiding centre radius $\Rg$, see Fig.~\ref{fig:sinfits}. It is natural to assume that the observed vertical oscillation links to the Monoceros ring and TriAnd overdensities in the outer disc, but detailed studies in the larger upcoming Gaia samples are required to confirm this. 

Using the proper motion of Sgr A$^\star$ to indicate the galactic standard of rest, there is a negative mean vertical motion at small $|\Lz|$, thus another deviation from the simple warp picture.

This pattern discovered in $\meanW$-$\Lz$ is stable against parallax accuracy and the size of the acceptance angle $\epsilon$. There is a tendency to more extreme $\meanW$ values for very large $|\Lz|$ when restricting to very small relative parallax errors, which also limits the sample to smaller distances. This is similar to a strong uptrend of $\meanW$ vs. $|\Lz|$ observed in the RAVE-TGAS sample, and originally in Hipparcos by \cite{Dehnen98}. The difference to the full sample is far larger than can be expected by distance errors. It is likely caused by local streaming motion in the Solar neighbourhood, although a mild kinematic bias by the kinematic selection in the Hipparcos sample (which gains weight at stricter astrometric quality cuts) is also possible. 

Inspection of the $V$-$W$ distribution does not indicate any streams relevant for this analysis (Fig.~\ref{fig:vwmap}). The only significant deviation in the $\meanW$-$\Lz$ relation from our naive sinusoidal fits in Fig.~\ref{fig:sinfits} is an upward spike in $\meanW$ near $-\Lz\sim2150 \kpc\kms$ (corresponding to a guiding centre of $9\kpc$) and is also visible in $\meanW$-$\Vphi$ plots at $|\Vphi| \sim 260 \kms$. The feature is far too narrow in $\Lz$ to belong to a bending mode and is likely a stream, which may even be mostly in the disc plane with the $W$ motion contaminated by large in-plane radial velocity. A remote possibility would be a resonance between stellar orbits and the galactic warp. 

More than showing for the first time at high significance both the stellar warp and vertical disc oscillations in local stellar kinematics, this study gives a first glimpse at the clarity with which Gaia will allow to map out even delicate structures in the Galactic disc.

We hope that these observations will trigger a new effort to a consistent theory of disc warps and vertical oscillations. We have here presented a very naive picture in which the mean vertical motions are linked to and mostly explained by the stellar angular momentum. This comes very close to the usual assumptions in classic modelling of disc waves and warps. However, resonant effects in this picture have to be explored in-depth. In this sample, most of our measurement relies on stars that are 1-2$\kpc$ away from their guiding centre radii. These stars cannot be treated as moving along tilted rings at their guiding centre, since they feel the vertical potential of the radius at which they are moving, where in the naive picture, they would be on average above or below the plane. \changed{Some resonant effects can be expected between the perturbation by the warp and the orbital motions.}. 

What we call ``warp'' throughout this paper, addressing the long-scale rise in $\meanW$ vs. $\Lz$ in contrast to the short-wavelength bending mode, should be interpreted carefully. We chose to use the word to match the previous literature, but due to the modest extent of our sample, one could equally choose to believe that this is just the shoulder of a much larger wave. How exactly both patterns link up to the outer disc oscillations remains to be explored in future datasets.

\section*{Acknowledgements}
\changed{We are grateful to the anonymous referee for a careful reading of the manuscript and useful comments.}
It is a pleasure to thank L. Eyer, M. Aumer, J. Binney, P. McMillan, E. d'Onghia, and C. Terquem for helpful discussions and comments. RS is supported by a Royal Society University Research Fellowship. WD is partly supported by STFC grant ST/N000757/1. This work used the DiRAC Data Centric system at Durham University, operated by the Institute for Computational Cosmology on behalf of the STFC DiRAC HPC Facility (www.dirac.ac.uk. This equipment was funded by a BIS National E-infrastructure capital grant ST/K00042X/1, STFC capital grant ST/K00087X/1, DiRAC Operations grant ST/K003267/1 and Durham University. DiRAC is part of the National E-Infrastructure. This work was supported by the European Research Council under the European Union's Seventh Framework Programme (FP7/2007-2013)/ERC grant agreement no.~321067. 
{This  work  has made use of data from the European Space Agency (ESA)  mission Gaia (http://www.cosmos.esa.int/gaia), processed by the Gaia Data Processing  and  Analysis  Consortium  (DPAC, http://www.cosmos.esa.int/web/gaia/dpac/consortium). Funding  for  the  DPAC  has  been  provided  by  national institutions, in particular the institutions participating in the Gaia Multilateral Agreement.}

\bibliographystyle{mnras}
\bibliography{paper}

\appendix

\section{Searching for substructure}

\begin{figure*}
\epsfig{file=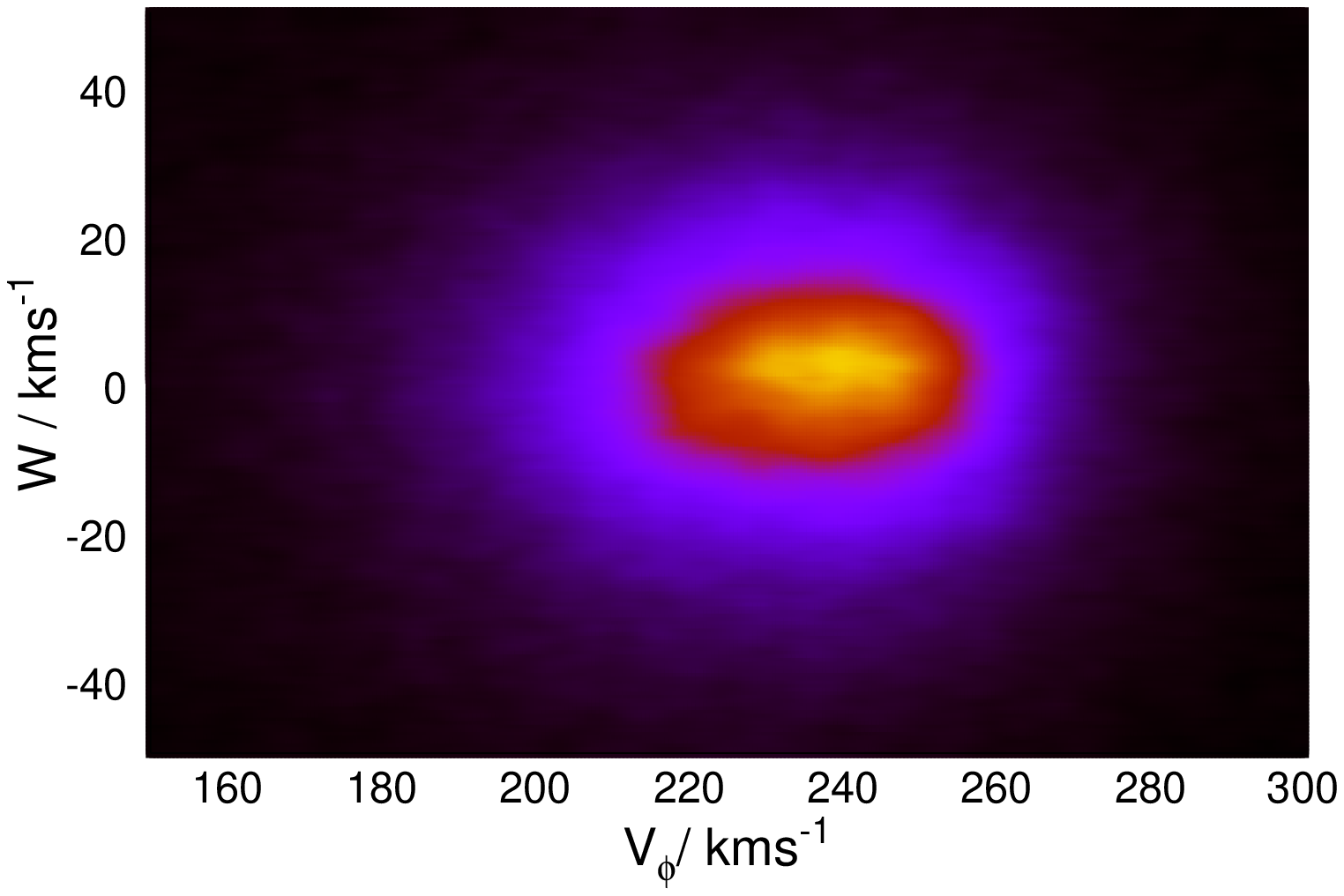,angle=-90,width=0.49\hsize}
\epsfig{file=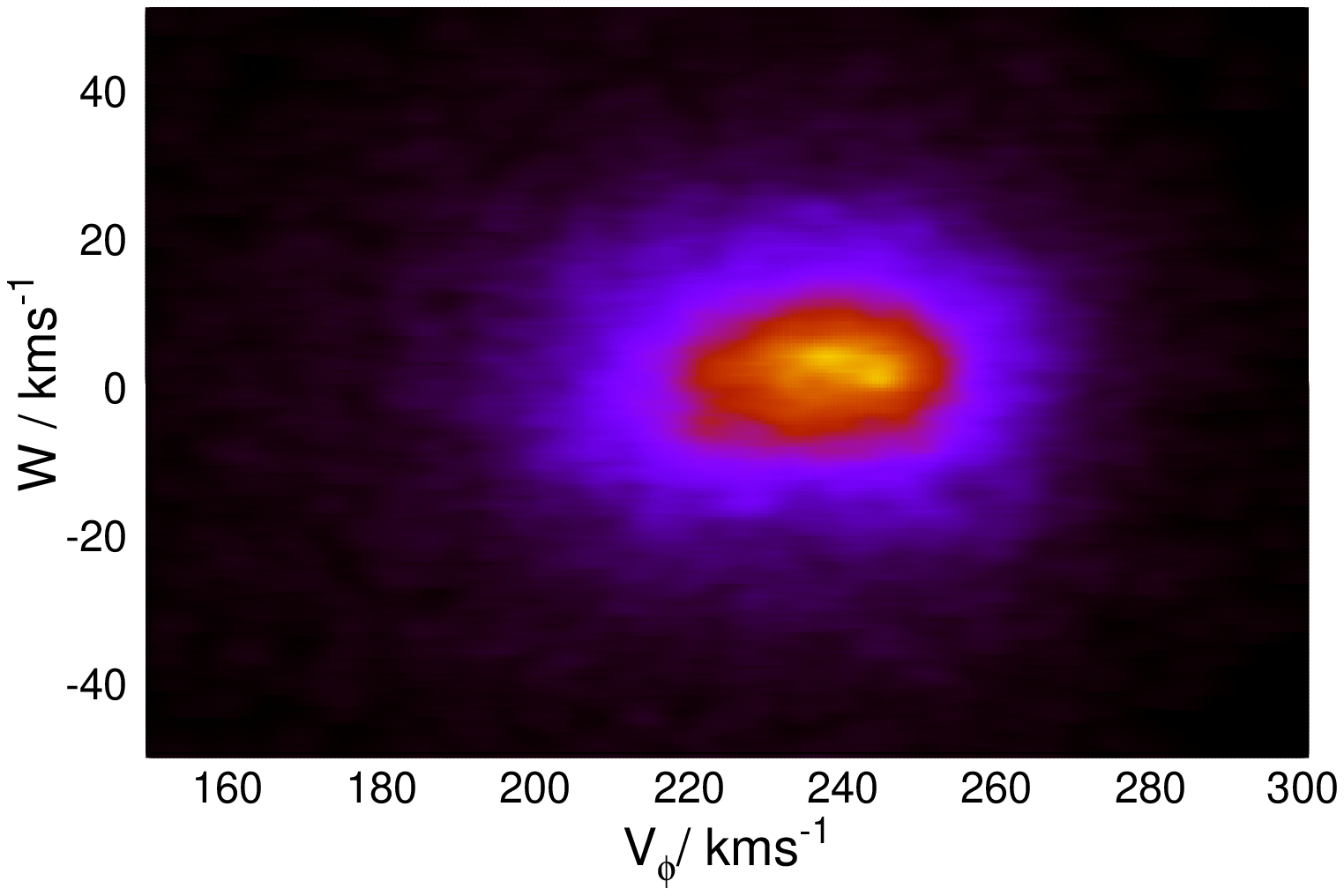,angle=-90,width=0.49\hsize}
\epsfig{file=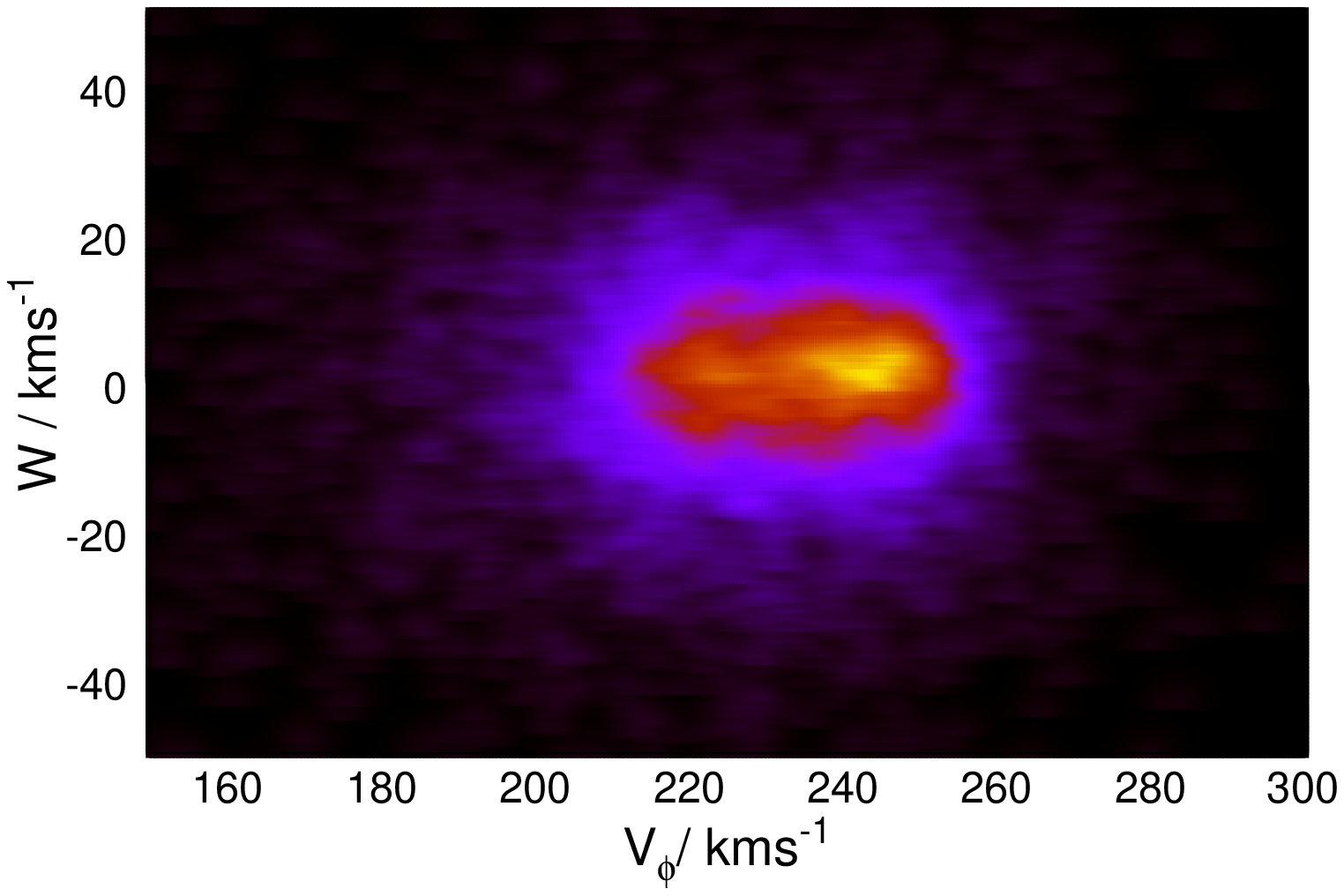,angle=-90,width=0.49\hsize}
\epsfig{file=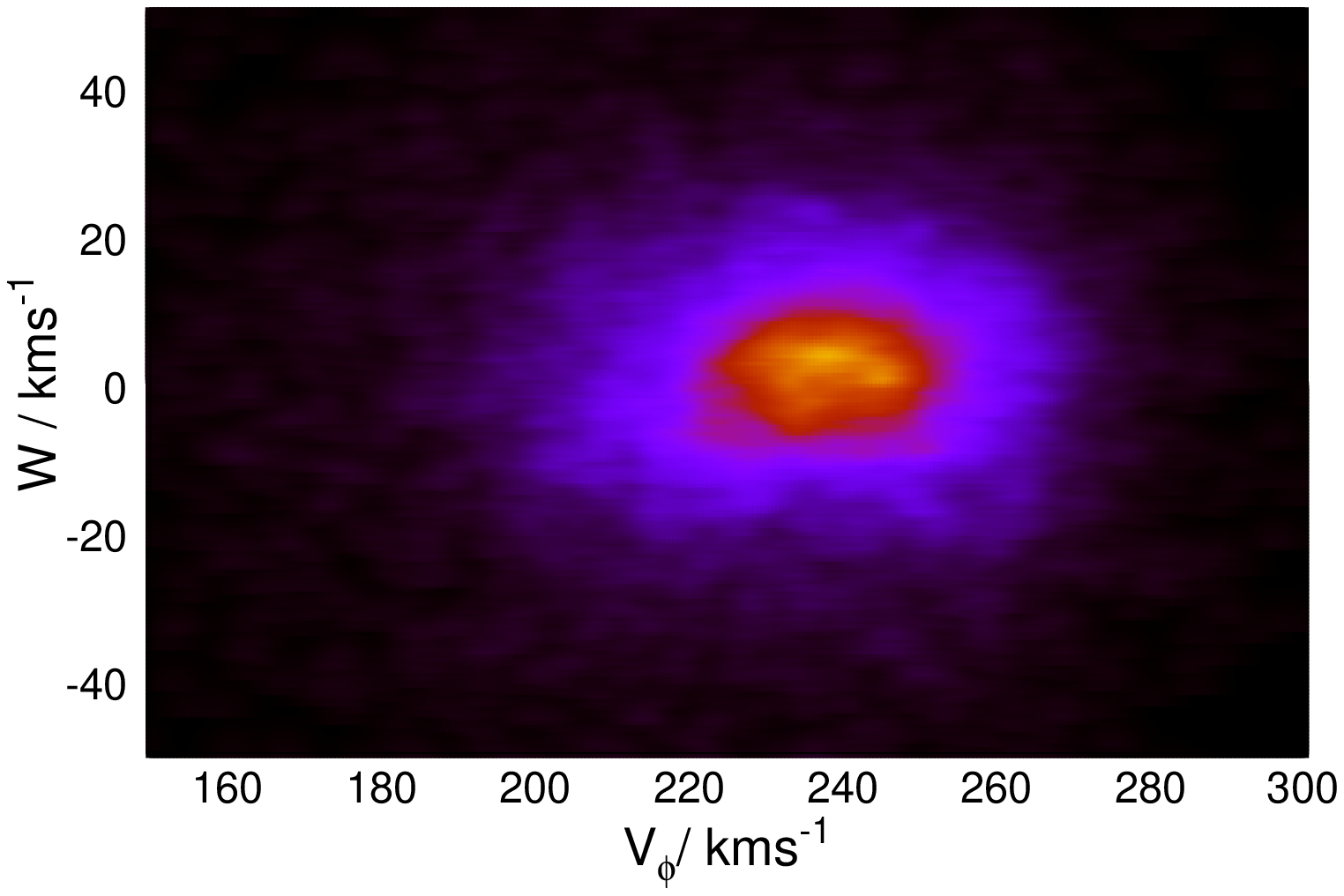,angle=-90,width=0.49\hsize}
\caption{Mapping the $\Vphi$-$W$ plane for the TGAS sample. We have folded the estimates for $\Vphi$ and $W$ from each star with a 2D Gaussian kernel of width $2\kms$. The panels show all stars within an acceptance angle $\epsilon=30 \degr$ (\emph{top left}), $15\degr$ (\emph{top right}), as well as stars in the outward (\emph{bottom left}) and inward (\emph{bottom right}) cones with $\epsilon=30 \degr$.}\label{fig:vwmap}
\end{figure*}

\begin{figure}
\epsfig{file=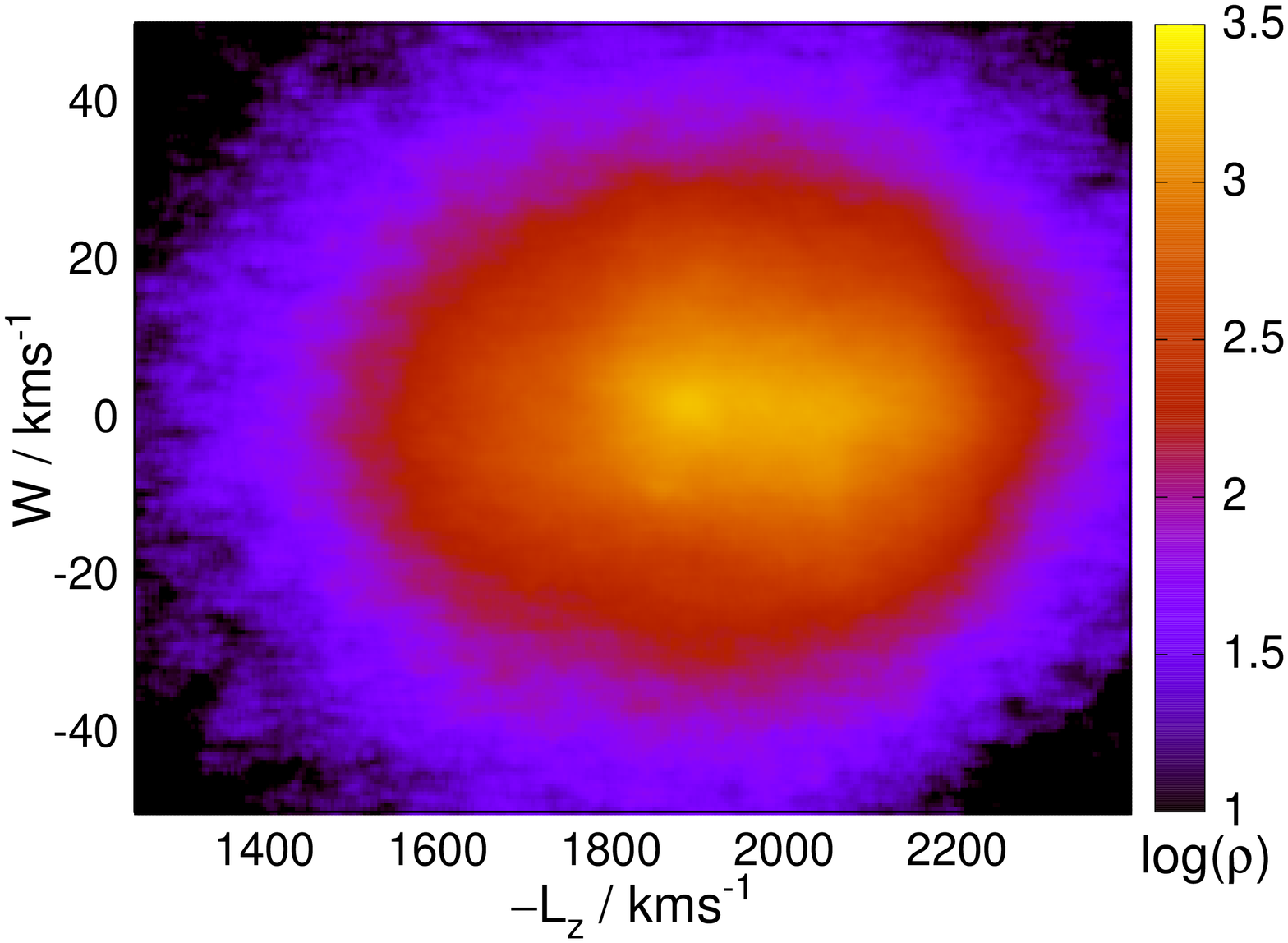,angle=-90,width=\hsize}
\caption{Logarithmic density in the $\Lz-W$ plane. We again used a 2D-Gaussian kernel with an equivalent of $2 \kms$ in $W$ and $\Vphi$ to reduce the shot noise.}\label{fig:sgauss}
\end{figure}

Dominant streams in the data could cause some localised deviations in $\meanW$. Now, we know that most disc streams have very small $\meanW$ velocities, and even if $10\%$ \changed{of the stars} in one angular momentum bin have a $\meanW$ of order $10 \kms$, they would merely bias that bin by $\sim 1 \kms$. In addition, streams or stream-like features tend to be very well-localised at a single $\Lz$ value, so they should be very narrow in $\Lz$. The only feature we could identify with some confidence this way, is the stream-like deviation around $-\Lz \approx 2140 \kpc \kms$. 

\section{Mock sample tests} \label{app:mock}

\begin{figure}
\epsfig{file=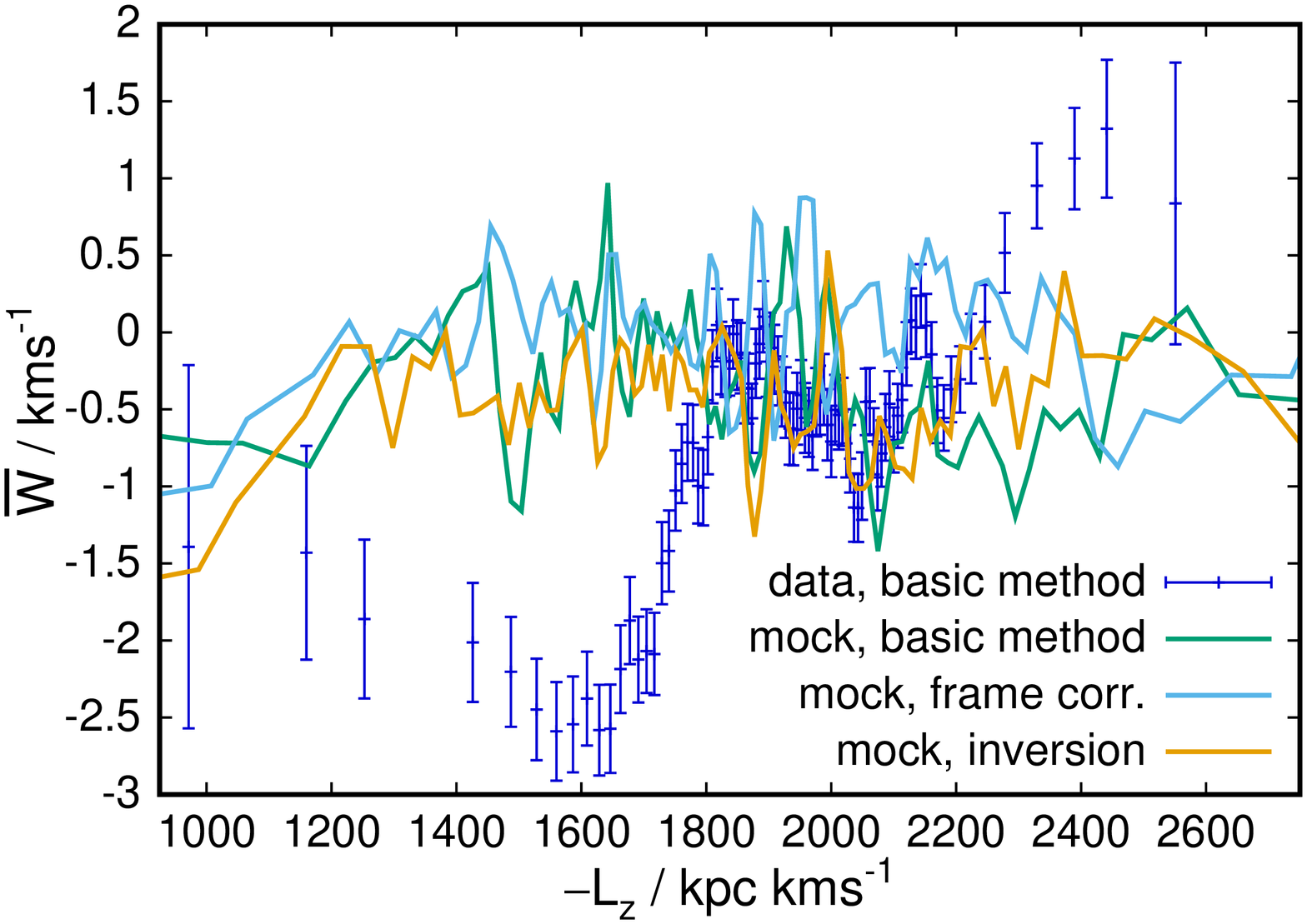,angle=-90,width=\hsize}
\caption{Mock sample tests for each method (lines) plotted together with the real result for the mock sample.}\label{fig:mock}
\end{figure}

\begin{figure}
 \epsfig{file=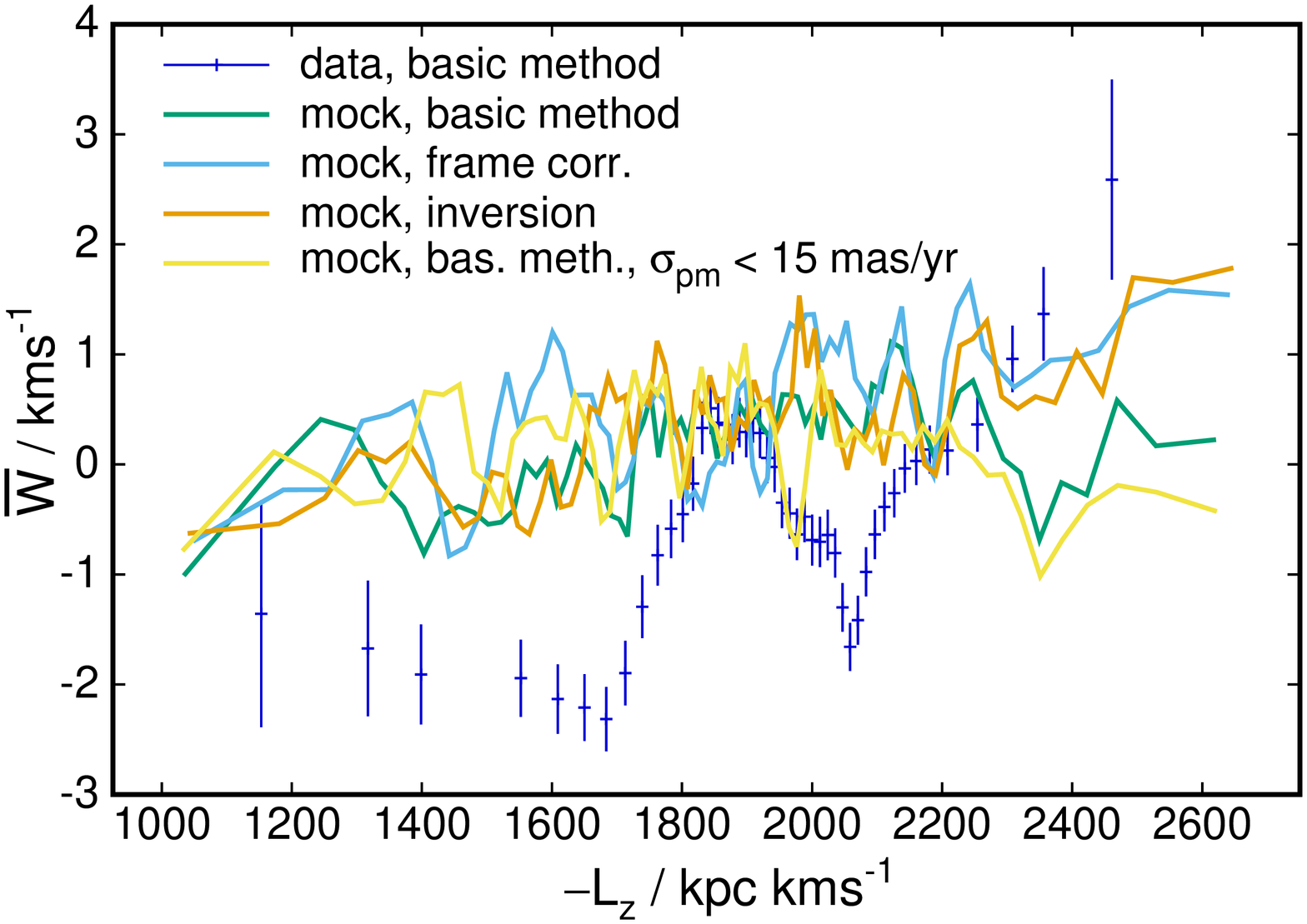,angle=-90,width=\hsize}
 \caption{\changed{Mock sample tests as in Fig.~\ref{fig:mock}, but now using the full error correlation matrix between $(p,\mura ,\mudec )$. The error bars depict the sample when cutting for $\sigma_{\mathrm{pm}} < 2 \mas \yr^{-1}$, and the additional line shows another mock sample for the frame correction method with $\sigma_{\mathrm{pm}} < 15 \mas \yr^{-1}$. The same pattern of trend + wave remains in the data with this quality cut, although the cut likely afflicts a kinematic bias by removing preferentially Tycho stars that are not in the Hipparcos set.}}\label{fig:mockmatrix}
\end{figure}

\begin{figure}
 \epsfig{file=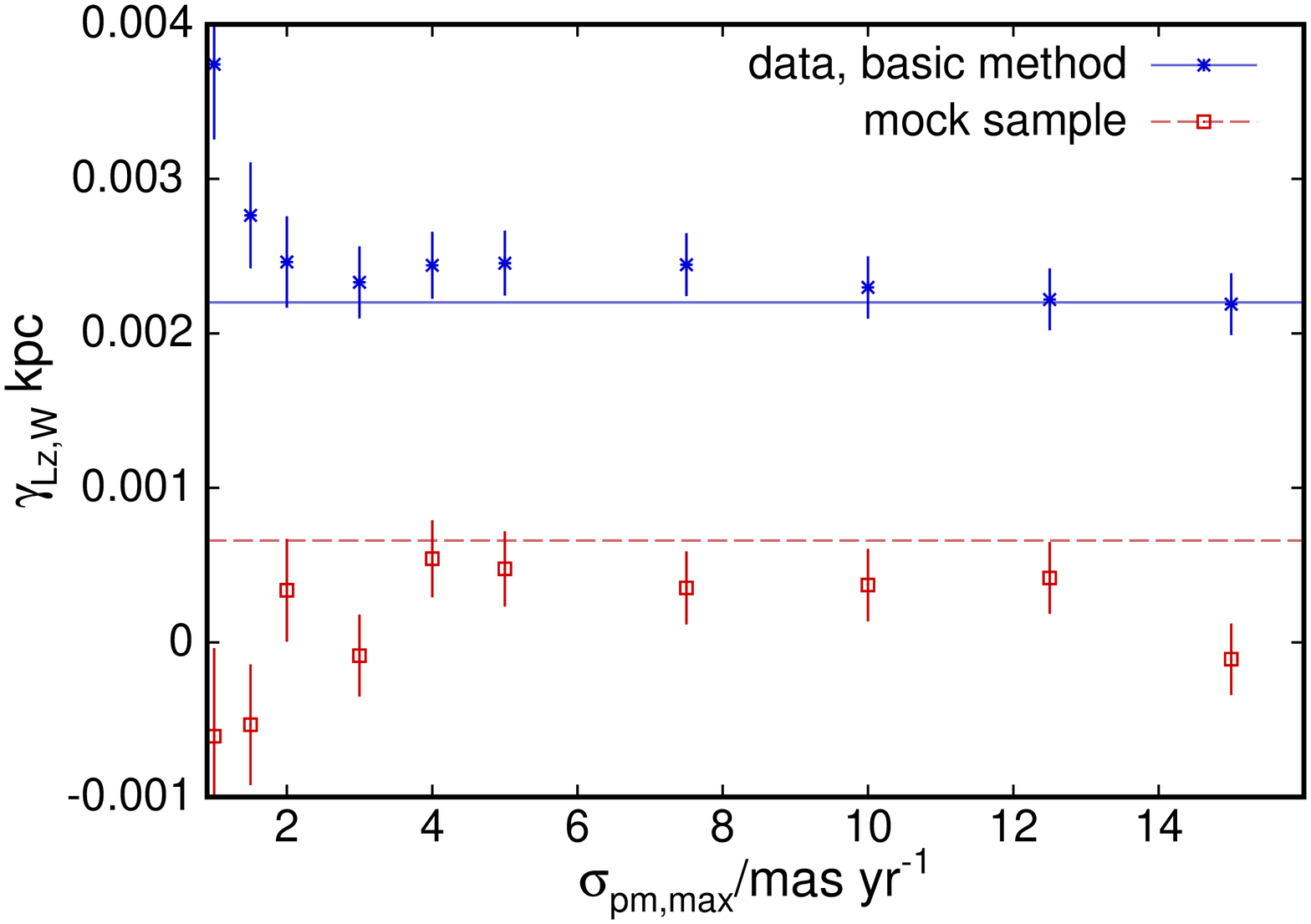,angle=-90,width=\hsize}
 \caption{\changed{Slope in a linear fit of $W$ vs. $\Lz$ when varying the proper motion quality cut $\sigma_{\mathrm{pm,max}}$ in the real vs. the mock sample. We note that this trend in the real sample is almost perfectly stable when varying the cut.}}\label{fig:mocktrend} 
\end{figure}

To check the correctness of our three different evaluation methods, we created mock samples for the Gaia-TGAS catalogue. For each star in the real sample, we took over the position and randomly drew a set of velocities, calculated the proper motions for the new set of velocities, and then re-estimated the velocity with each method. For the sake of simplicity, the velocities are drawn from two Gaussian distributions, which have a mean azimuthal velocity of $225 \kms$ and a velocity ellipsoid with $(\sigma_U, \sigma_V, \sigma_W) = (50, 40, 30) \kms$, and with $0.01\%$ likelihood a zero mean velocity and $(\sigma_U, \sigma_V, \sigma_W) = (140, 90, 90) \kms$. 

The results of these mock sample tests are shown in Fig.~\ref{fig:mock}. As expected, the mock samples neither show any appreciable spurious trend, nor any pattern that remotely resembles the signal we measure in the data. To enhance any possible biases, we have chosen an unreasonably large velocity dispersion in the mock samples, which leads to a bit enhanced scatter in each single value. This way, the figure also demonstrates how much larger the measured signal is compared to noise expected for the sample. As a second test, we also varied the parallaxes for the samples, but the resulting distance biases only lead to a very minor (of order $0.1 \kms$) tendency of both the high- and low-$\Lz$ tails to have slightly slower $\meanW$ than the centre of the distributions, since distance overestimates tend to accumulate in the tails.

\changed{Another important consideration is the correlation between distance and $\RA$/$\DEC$ proper motion errors in Gaia-TGAS. Since the cones in which we observe have a relatively small opening angle, and since the correlation matrix in Gaia-TGAS varies systematically and quite smoothly with position on the sky, there could be a residual effect on the sample. However, given the agreement between the inward and outward cones in Fig.~\ref{fig:TgasVW}, it is already highly unlikely that the observed wave-pattern is created by measurement uncertainties, but it would in principle be possible that errors contribute to the general trend of $\meanW$ vs. $\Lz$. To examine this possibility we created mock samples as in Fig.~\ref{fig:mock}, and draw the mock measurements from the full covariance matrix, i.e.\ incorporating the reported correlations between $(p, \mura, \mudec)$. 

The result is shown in Fig.~\ref{fig:mockmatrix}. To dispel any concerns that the error correlations could have anything to do with the observed wave pattern, we now show the actual data with error bars for all stars with a total proper motion error $\sigma_{\mathrm{pm}} \def \sqrt{\sigma_{\mathrm{RA}}^2 + \sigma_{\mathrm{DEC}}^2} < 2 \mas \yr^{-1}$. While this subset should not be used due to the contamination with the Hipparcos selection function (Hipparcos stars have better proper motion measurements, but are kinematically selected), we can see that the general structure does not change. In the mock samples, two of the three full samples show a very slight uptrend, so we also show the equivalent mock catalogue for a very generous cut of $\sigma_{\mathrm{pm}} < 15 \mas \yr^{-1}$, which does not have any trend. 

Fig.~\ref{fig:mocktrend} shows an evaluation of the trend $\gamma_{L_z,W}$ between $\meanW$ and $\Lz$, using the basic method, i.e. the slope in a linear fit of $W$ vs. $\Lz$ for samples with different proper motion quality cuts $\sigma_{\mathrm{pm,max}}$, both for mock samples and the data. The horizontal line shows the trend in the full sample, the dashed line depicts the trend in the full mock sample. While the slight trend in the full mock sample could have been taken as a suggestion that a very small amount of the trend in the data might stem from correlated proper motion errors, there is no decrease in the observed trend when we start applying cuts to the data in $\sigma_{\mathrm{pm}}$. The stronger deviations at $\sigma_{\mathrm{pm,max}} < 2 \mas \yr^{-1}$ in the data are not consistent with trends in the mock catalogues, and likely result from a lucky sample selection at this small sample size or the increasing importance of the Hipparcos selection function.

Distance errors could also provide a very small contribution: while most of the usual baseline effect \citep[distance overestimates carry larger weight due to their position on the x-axis away from the mean, see][]{SBA} is compensated because distance overestimates have a negative $W$ bias both for large and small $\Lz$, there could be a minor effect due to the different base lengths (in $\Vphi - \Vsun$) on both sides. However, no significant trend is detected either, with the mean mock sample trend $\gamma_{L_z,W,\mathrm{mock}} = (0.00011 \pm 0.00013) \mas \yr^{-1}$. This level is about a factor $20$ below our detection.
}

\label{lastpage}
\end{document}